\documentclass[12pt]{article}

\newcommand{\nc}{\newcommand}
\nc{\beq}{\begin{equation}}
\nc{\eeq}{\end{equation}}
\nc{\beqa}{\begin{eqnarray}}
\nc{\eeqa}{\end{eqnarray}}

\input epsf
\newwrite\ffile\global\newcount\figno \global\figno=1

\def\writedef#1{}

\def\figin{\epsfcheck\figin}\def\figins{\epsfcheck\figins}
\def\epsfcheck{\ifx\epsfbox\UnDeFiNeD
\message{(NO epsf.tex, FIGURES WILL BE IGNORED)}
\gdef\figin##1{\vskip2in}\gdef\figins##1{\hskip.5in}% blank space instead
\else\message{(FIGURES WILL BE INCLUDED)}%
\gdef\figin##1{##1}\gdef\figins##1{##1}\fi}

\def\figinsert{}
\def\ifig#1#2#3{\xdef#1{fig.~\the\figno}
\writedef{#1\leftbracket fig.\noexpand~\the\figno}%
\figinsert\figin{\centerline{#3}}\medskip\centerline{\vbox{\baselineskip12pt
\advance\hsize by -1truein\center\footnotesize{  Fig.~\the\figno.} #2}}
\bigskip\endinsert\global\advance\figno by1}
\def\endinsert{}

\begin{document}

\title{\large{\bf Bounds on Flavor Gauge Bosons from precision Electroweak 
Data}}

\author{
Gustavo Burdman\thanks{burdman@pheno.physics.wisc.edu} \\
{\small\em Department of Physics, University of Wisconsin, Madison, 
WI 53706, USA.} \\ 
R. Sekhar Chivukula\thanks{sekhar@bu.edu} \\
{\small\em Department of Physics,
Boston University, Boston, MA 02215, USA.} \\ 
Nick Evans\thanks{n.evans@phys.soton.ac.uk}
\\ {\small\em Department of Physics,
University of Southampton, Southampton, S017 1BJ, UK.} }

\date{June 1999}

\maketitle

\begin{picture}(0,0)(0,0)
\put(350,335){BUHEP-99-12}
\put(350,320){MADPH-99-1120}
\put(350,305){SHEP-99-05}
%\put(350,290){****DRAFT****}
\end{picture}

\begin{abstract}
  Gauged flavor symmetries at low energies have been proposed in models
  of dynamical electroweak symmetry breaking and fermion mass
  generation. The massive flavor gauge bosons give rise to corrections
  to precisely measured electroweak quantities. We perform a fit to the
  collider  electroweak data and place indirect limits on such new
  physics. In particular we study several models from the literature:
  universal coloron; chiral top color; chiral quark family symmetry;
  SU(9) and SU(12) chiral flavor symmetry. The $95\%$ exclusion limits
  on the mass of the heavy gauge bosons for these models, at their
  critical coupling for chiral symmetry breaking, typically lie between
  1-3 TeV. We discuss the robustness of these bounds with respect to
  changes in higgs mass, b-quark asymmetry data, the SLD left right 
  asymmetry data,  or additional new physics.
\end{abstract}

\newpage

\section{Introduction}

The Standard Model (SM), in the limit where we neglect all gauge
interactions and fermion masses, has a large SU(45) global symmetry
corresponding to the fact that in this limit there are 45 chiral fermion
fields that are indistinguishable. The gauge interactions of the SM are
by necessity subgroups of this maximal symmetry, they are in particular
the familiar $SU(3)_c \times SU(2)_L \times U(1)_Y$ gauge symmetry.
There is no current theoretical understanding of why this particular
subgroup is picked out to be gauged. Furthermore we have learned from the
SM that gauge symmetries may be broken and not manifest at low energies.
The possibility therefore exists that at high energies a larger subgroup
of the SU(45) symmetry is gauged, and then broken by some dynamics or higgs
mechanism to the SM gauge group.

These gauged flavor symmetries have been invoked in a number of scenarios 
to play a role in the dynamical generation of fermion masses. For example
they may play the part of extended technicolor \cite{ETC}
interactions in technicolor
models \cite{TC} or top condensation models \cite{tc}, 
feeding the electroweak symmetry (EWS) 
breaking fermion condensate down to provide masses for the lighter standard
model fermions. Strongly interacting flavor 
gauge interactions may also be 
responsible for the condensation of the fermions directly involved in
EWS breaking. For example, top condensation has been postulated to result from
a topcolor gauge group \cite{tcol} and, in the model of 
ref. \cite{king}, from family gauge interactions. There has been renewed interest in
these models recently with the realization that variants, in which the 
top mixes with singlet quarks, can give rise to both the EW scale and an
acceptable top mass via a seesaw mass spectrum \cite{tseesaw}. These
top seesaw models have the added benefit of a decoupling limit which
allows the presence of the singlet fields to be suppressed in precision EW
measurements, bringing these dynamical models 
in line with the data. Flavor universal 
variants \cite{fuseesaw} with the dynamics 
driven by family or large flavor gauge symmetries
have also been constructed.

The naive gauging of flavor symmetries at low scales (of order a few
TeV) often gives rise to unacceptably large flavor changing neutral
currents (FCNC) since gauge and mass eigenstates do not coincide (gauge
symmetries that give rise to direct contributions to $K^0-\bar{K}^0$
mixing typically are constrained to lie above 500 TeV in mass scale).
There are, though, many models that survive these constraints.  Gauge
groups that only act on the third family are less experimentally
constrained - topcolor is such an example. Models in which the chiral
flavor symmetries of the SM fermions are gauged preserving the SM
$U(3)^5$ flavor symmetry \cite{ctsm} respect the GIM mechanism even
above the breaking scale and do not give rise to tree level FCNCs
\cite{georgi}.  There are also strong constraints on gauged flavor
models where the dynamics responsible for the breaking of the flavor
symmetry does not respect custodial isospin \cite{mixing}.  We shall
restrict ourselves to models where the top mass is the sole source of
custodial isospin breaking.

Since these interactions may exist at relatively low scales (a few TeV),
where they do not completely decouple, and play an integral part in
either EWS breaking or fermion mass generation, it is interesting to
study the current experimental bounds.  In this paper we will
concentrate on the bounds from precision EW measurements \cite{data}. We
will constrain a number of models that have been proposed in this
context: the universal coloron model \cite{unicol}; chiral top color
\cite{tseesaw}; chiral quark family gauge symmetry
\cite{georgi,king,fuseesaw}; SU(9) chiral quark flavor symmetry
\cite{randall,fuseesaw}; and SU(12) chiral, flavor symmetry
\cite{randall,fuseesaw}.  Except in the last section, for simplicity the
only fermions included in our analysis are the ordinary quarks and
leptons. Therefore, as presented here the models will not be anomaly
free. The models can be made anomaly free by the introduction of
additional fermions that obtain masses at or above the gauge boson mass
scale as a result of the flavor symmetry breaking sector. The SM
fermions that are light below that scale survive because they are chiral
under the remaining, unbroken gauge interactions - this is an example of
Georgi's ``extended survival hypothesis'' \cite{survival}.  For each of
the gauge symmetries we study, an explicit example of an anomaly-free
model exists in the literature in the references provided above.

The gauge bosons can enter into precision EW predictions either through
corrections to the $Z$ mass from isospin violating 
effects in top loops \cite{toploop}
and mixing with the $Z$ \cite{mixing}, or as 
vertex corrections to the $Z$-fermion vertices \cite{vertex}.
We review the generic form of these corrections in Section 2. In Section 3
we describe our fit to the $Z$-pole data. In Section 4 we review each of
the models under study and display the precision limits. To this point the 
limits are on the models with only the flavor gauge bosons as new physics.
The contributions to the $\rho$ parameter are a significant
constraint and our limits correspond to  the hard limit
on $\Delta \rho$ (or $\alpha T$) of 
$\leq 0.1\%$ \cite{pdg}. The softer upper bound on $\Delta \rho$ 
often quoted of $0.4\%$ \cite{pdg}
assumes the existence of extra new physics contributing to the S parameter 
\cite{peskin}
to maximize the possible value of $\Delta \rho$. The flavorons  do 
not contribute to S (at leading order) and so the harder limit is applicable 
here. In Section 5 we discuss the dependence of the limits on changes
to the higgs mass. The limits are largely insensitive except in the case
of models which give rise to large corrections to the 
$\Delta \rho$/T parameter (in particular the coloron models) and prefer a
heavy higgs in order to cancel that contribution. The bounds on these
models decrease for higgs masses as large as $650$ GeV  which, without the 
flavorons, would be ruled out at $95\%$ confidence in the SM. We also discuss
the dependence of the limits on the b quark asymmetries which are the 
measurements with the most deviation from the SM predictions, and the SLD 
measurement of the left right asymmetry which provides one of the strongest
constraints on the higgs mass.

It is important to remember the limitation of indirect constraints such
as these; additional unanticipated new physics may exist that enters the
fit and in principle may cancel all contributions from the gauge bosons,
completely removing any bound (or equally the bound may be considerably
toughened by extra new physics)!  In practice such a conspiracy seems
unlikely and the bounds we calculate provide a sensible guide to the
exclusion limit. As an example of the possible effect of additional new
physics, in Section 5 we also discuss the addition of vertex corrections
from mixing between the SM fermions and singlet massive fermions such as
occurs in the top \cite{tseesaw} and flavor universal \cite{fuseesaw}
see-saw models. Another possible source of S and T contributions are the
scalar sectors of these models, which are potentially light. However,
enumerating these scalars is model dependent and their masses are hard
to estimate reliably so we do not attempt to include them. We expect the
few TeV bound on the flavoron masses coming from our fit to hold even
with such perturbations.

\section{Flavoron Corrections to EW Parameters}

The massive flavor gauge bosons (flavorons) under study give a number of
corrections to low energy EW parameters. In this section we provide
generic calculations of these effects. 

$\left. \right.$   \hspace{-0.2in}\ifig\prtbdiag{}
{\epsfxsize5truecm\epsfbox{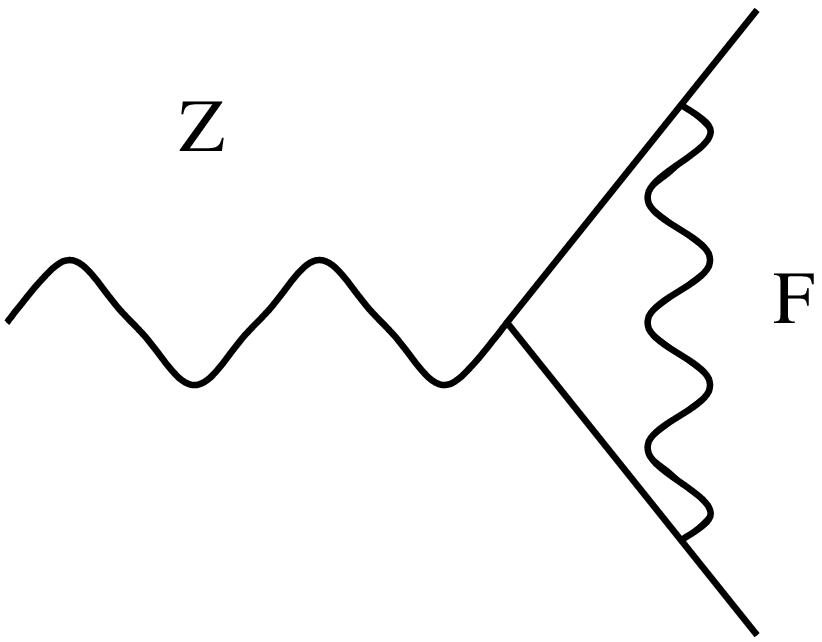}} \vspace{-1cm}

\begin{center}
Figure 1: $Z$ vertex corrections from heavy flavorons. 
\end{center}

Flavorons  acting on the SM fermions give rise to 
vertex corrections in the low energy SM. We assume that there are
no massive fermions beyond those in the SM transforming under the 
flavor gauge groups as part of the same gauge multiplet as the SM
fermions (an example would be techni-fermions). These would
give rise to overly large corrections, as is well known from 
extended technicolor models \cite{Zbb}.
The main corrections to $Z$ vertices occur at order $p^2$ in the expansion of
the penguin vertex correction in Fig 1. Hill and Zhang \cite{vertex}
have calculated 
the correction to the vertex to be
\beq \label{vertex}
\Delta g_f = g_f {G \kappa_F  \over 6 \pi}   {M_Z^2 \over M_F^2}
\ln \left({M_F^2 \over M_Z^2} \right)
\eeq
where $g_f$ is the coupling for the $Z$-fermion vertex,
$G$ is the appropriate group theory factor for the diagram, 
$\kappa_F = g_F^2/4 \pi$ with $g_F$ the
flavoron gauge coupling, and $M_F$ the flavoron mass.

$\left. \right.$   \hspace{-0.2in}\ifig\prtbdiag{}
{\epsfxsize7truecm\epsfbox{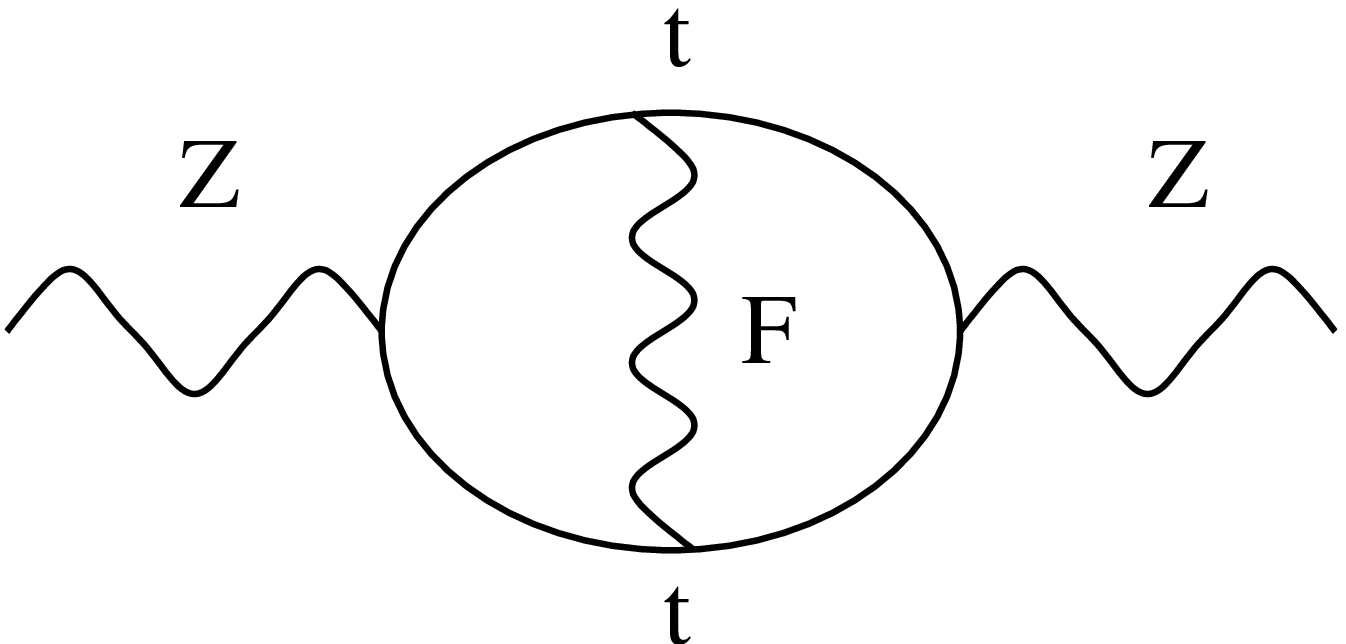}}  \vspace{-1cm}

\begin{center}
Figure 2: Flavoron corrections to the $Z$ mass through top loops. \vspace{1cm}
\end{center}

The flavorons can also give corrections to the $Z$ mass through the diagram 
in Fig 2 involving a top loop \cite{toploop}. 
The top bottom mass splitting provides the
source of isospin breaking so there are contributions to 
$\Delta \rho$ ($T$) even when the flavoron
physics is isospin preserving. 

The contribution to T is
\beq
T = {\pi \over s^2_{\theta_w} c^2_{\theta_w} M_Z^2} \left[ 2 \Pi_{LL}(m_t,0)
- \Pi_{LL}(m_t,m_t) \right]
\eeq
where $\Pi_{LL}(m_t,0)$ and $\Pi_{LL}(m_t,m_t)$ refer to the left-handed 
vacuum polarizations involving one top and one (massless) bottom quark,
and two top quarks in the loop respectively, and they are evaluated at 
zero momentum transfer. Following \cite{toploop} we will approximate the
two loop diagram by a product of two one loop diagrams after the 
flavoron propagator has been contracted to a point interaction. 

Allowing the flavoron to have different couplings to left and right
handed top quarks we find
\begin{eqnarray}
\Pi_{LL}(m_t,0) & = & -{N_c G'_{LL} \over 64 \pi^3} m_t^4
\left[ \ln ( M_F^2 /m_t^2 ) \right]^2 {\kappa_F \over M_F^2}\\
\Pi_{LL}(m_t,m_t) & = & -{N_c (G'_{LL} + G'_{RR}) \over 16 \pi^3} m_t^4
\left[ \ln ( M_F^2 /m_t^2 ) \right]^2 {\kappa_F \over M_F^2}
\end{eqnarray}
with $G'_{LL}$ and $G'_{RR}$ appropriate group theory factors for the model
for the interaction between two left handed quarks and 
two right handed quarks respectively, and where we have evaluated the 
leading logarithm. The resulting 
contribution to T is
\beq \label{tloop}
T = N_c ( G'_{LL} + 2G'_{RR}) 
{m_t^4  \over 32 \pi^2 s_{\theta_w}^2 c_{\theta_w}^2
M_Z^2}  {\kappa_F \over M_F^2} \left[ \ln(M_F^2/m_t^2) \right]^2
\eeq

$\left. \right.$   \hspace{-0.2in}\ifig\prtbdiag{}
{\epsfxsize7truecm\epsfbox{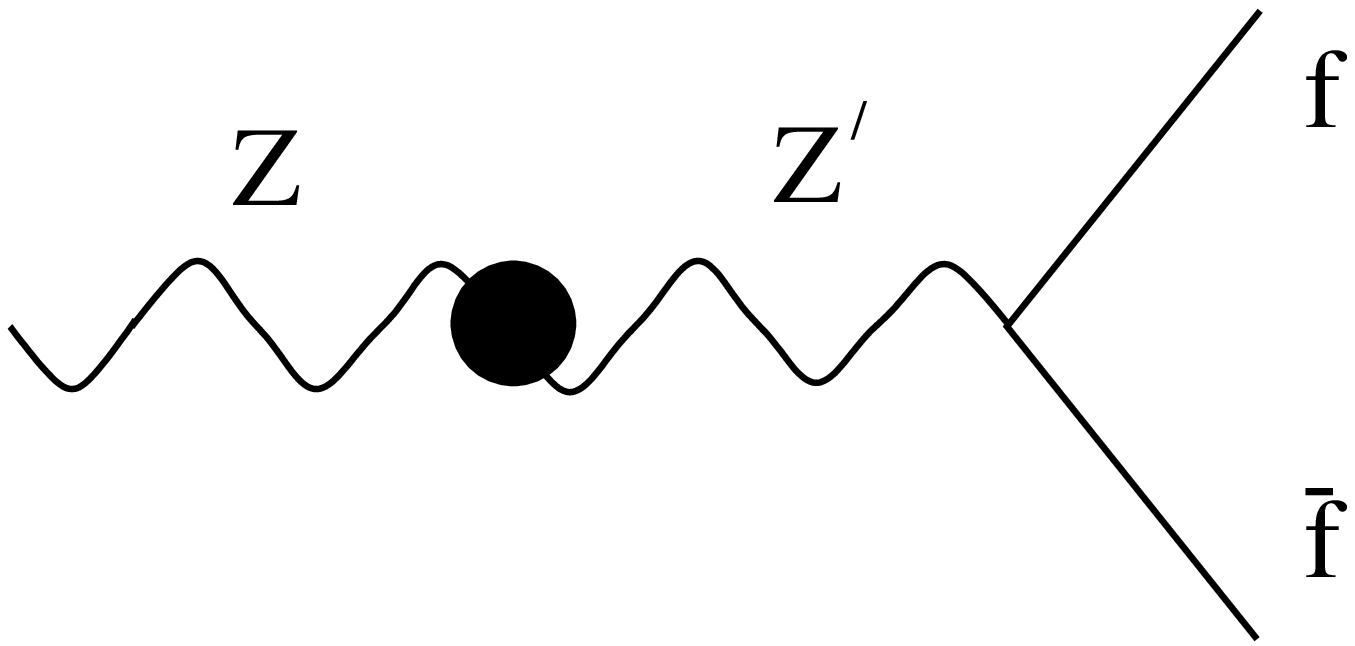}} \vspace{-1cm}

\begin{center}
Figure 3: Flavoron-$Z$ mixing shift to SM couplings. 
\end{center}

There is a final possible source of contributions to the $Z$ vertex corrections
and the $T$ parameter in models 
where a generator of the flavor group (we will call the associated 
gauge boson $Z'$) mixes with hypercharge or $T_3$ of
the weak SU(2) group \cite{mixing}. 
If at the weak scale the $Z$ $Z'$ mass matrix takes the form
\beq
(Z^\mu Z'^\mu) \left( \begin{array}{cc} M_Z^2 &  \Delta M^2 \\ \Delta M^2 &
M_F^2 \end{array} \right) \left( \begin{array}{c} Z_\mu \\ Z'_\mu \end{array}
\right)~,
\eeq
and if $M^2_F \gg M^2_Z,\, \Delta M^2$
then there will be corrections to the $Z$-fermion couplings in the low energy 
theory of the form (from Fig 3)
\beq
\Delta g_{f} \simeq - {\Delta M^2 \over M_F^2} g^{Z'}_f
\eeq
and, diagonalizing the mass matrix so that the shift in the low energy 
$Z$ mass is
\beq
\Delta M_Z^2 = -\,{(\Delta M^2)^2 \over M_F^2}
\eeq
gives a T parameter contribution
\beq
T = +\,{\Delta M_Z^2 \over \alpha M_Z^2}
\eeq
Clearly for these contributions to be small we require that the only
sources of custodial isospin breaking occur at the weak scale so $\Delta
M$ is of order $v$ and not of order $M_F$. The models we consider all
respect custodial isospin at the flavor symmetry breaking scale.

These shifts in low energy SM parameters  are the leading corrections. The
shifts in the $Z$-fermion vertices will for example also give rise to
S parameter contributions at loop level \cite{terningS} 
but these are considerably smaller
than the direct effects and we neglect them. 

\section{Fitting To Precision Data}

To place bounds on the models we use primarily the $Z$ pole precision 
observables from the LEP and SLD experiments plus the collider 
measurements of the W mass. 
We use the data and SM fit from \cite{data}

\beq \label{Vi}
\begin{array}{c|ccc}
V_i & {\rm Tree ~ Level} & {\rm Measurement} & {\rm SM ~fit ~pull} \\
\hline
\Gamma_Z(GeV) &  & 2.4939 \pm 0.0024 & -0.80 \\
\sigma_0 & \Gamma_{lep} \Gamma_{had} / \Gamma_{Z} & 41.491 
\pm 0.058 & 0.31\\
R_l &  \Gamma_{had} / \Gamma_{lep}& 20.765 \pm 0.026 & 0.66\\
R_b &  N_c (g_{b_L}^2 + g_{b_R}^2) / \Gamma_{had} 
& 0.21656 \pm 0.00074 & 0.90 \\
R_c  & N_c  (g_{c_L}^2 + g_{c_R}^2) / \Gamma_{had}
& 0.1735 \pm 0.0044 & 0.29\\
A_e & (g_{e_L}^2 - g_{e_R}^2) /(g_{e_L}^2 + g_{e_R}^2)
& 0.1479 \pm 0.0051 & 0.25\\
A_b & (g_{b_L}^2 - g_{b_R}^2) /(g_{b_L}^2 + g_{b_R}^2) 
& 0.867 \pm 0.035 & -1.93\\
A_c & (g_{c_L}^2 - g_{c_R}^2) /(g_{c_L}^2 + g_{c_R}^2) 
& 0.647 \pm 0.040 & -0.52\\
A^{0,b}_{fb} & 3 A_e A_b /4  & 0.0990 \pm 0.0021 & -1.81\\
A^{0,c}_{fb} & 3 A_e A_c /4& 0.0709 \pm 0.0044 & -0.58\\
A^{0,e}_{fb} & 3A_e^2/4 & 0.01683 \pm 0.00096 & 0.73\\
A_\tau &(g_{\tau_L}^2 - g_{\tau_R}^2) /(g_{\tau_L}^2 + g_{\tau_R}^2) 
& 0.1431 \pm 0.0045 & -0.79 \\
A_{LR} & A_e & 0.1510 \pm 0.0025 & 1.75 \\
M_{W}^{LEP2} & M^{SM}_{W} (1 + 0.0055\,T -0.0036\,S) & 80.37 \pm 0.09 & -0.01 \\
M_{W}^{pp} & M^{SM}_{W} (1 + 0.0055\,T -0.0036\,S) & 80.41 \pm 0.09 & 0.43
\end{array}
\eeq
where
\beq \begin{array}{c}
\Gamma_{Z} = \sum_f (g_{f_L}^2 + g_{f_R}^2)\\
\Gamma_{had} = \sum_{had} (g_{h_L}^2 + g_{h_R}^2)\\
\Gamma_{lep} = \sum_{lep} (g_{l_L}^2 + g_{l_R}^2)
\end{array} \eeq
The top quark, of course, does not contribute to any Z-pole variables at 
tree level. 

We use the fit of \cite{data} as our test Standard Model into which we
introduce corrections \cite{london}.  The full fit of \cite{data}
includes measurements of the Z mass and $\sin^2\theta_W$ as determined
in low energy neutrino scattering data.  For simplicity, in comparing
the predictions of our models to electroweak data we use the Z mass
(the most precisely measured quantity) as a (fixed) input to our
calculation, and do not include $\sin^2\theta_W$. 

As an illustration of the correspondence of our fit to that given in
\cite{data}, we may compare the constraints on the higgs mass in the
standard model.  The fit of \cite{data} corresponds to a best fit
higgs mass of $m_h = 76$ GeV, and a bound of 262 GeV at 95\%
confidence level.  Variation in the higgs mass may be viewed as
contributions to the S and T parameters which are calculated in
\cite{hagiwara}.  Using this method to incorporate variations in the
higgs mass and with the more limited set of precision variables we
consider, we find a $95\%$ confidence limit upper bound of $230$ GeV
($120$ GeV at $67\%$), in good agreement with the full fit.

Furthermore, because the higgs mass enters precision measurements only
logarithmically, the limit on the higgs mass is very sensitive to small
changes in the measured parameters and the difference in the higgs
mass bound exaggerates the difference between the two fits. In
contrast, the flavoron masses enter quadratically in corrections to
precision electroweak observables, and the bounds we derive would not
change significantly were we to incorporate $M_Z$ and
$\sin^2\theta_W$.  For definiteness, we will take a higgs mass of 100
GeV as our standard in the fits of section 4.  In Section 5 we show
that the bounds are largely insensitive to changes in the higgs mass.

%S and T are
%incorporated into the EW variables in (\ref{Vi}) through \cite{london}
%
%\beq \label{ST}
%\begin{array}{l}
%\Gamma_Z = \Gamma_Z^{SM} (1 - 0.00385  S + 0.0105 T) \\
%\Gamma_{had} = \Gamma_{had}^{SM}(1 - 0.00518  S + 0.0114 T)\\
%\Gamma_{lep} = \Gamma_{lep}^{SM}(1 - 0.00230  S + 0.00944 T)\\
%A_e = A_\tau = A_e^{SM} - 0.0284 S + 0.0201 T\\
%A_b = A_b^{SM}  - 0.0066 S + ??? T\\
%A_c = A_c^{SM} + 0.0179 S + ??? T\\
%R_b = R_b^{SM}(1 + 0.00066  S - 0.0004 T)\\
%R_c = R_c^{SM}(1 - 0.00131  S + ???? T)
%\end{array}
%\eeq
%
%The T parameter enters as an effective shift in 
%the calculated value of $\sin^2 \theta_w$.  The value of $\sin^2 \theta_w$ 
%is extracted from the relation
%
%\beq
%s_{\theta_w}^2 c_{\theta_w}^2 = {\pi \alpha \over \sqrt{2} G_F M_Z^2 (1+\delta% \rho)}~,
%\eeq
%from which it follows that T enters 
%
%\beq \label{Tssq}
%\delta s^2_{\theta_w} = - { \alpha s_{\theta_w}^2 c_{\theta_w}^2 
%\over 1-2s_{\theta_w}} T
%\eeq
%
%Note that the $Z$ pole variables are all ratios of couplings except for 
%$\Gamma_{Z}$, so only the latter is affected by shifts in the precursor
%$e / s_{\theta_w} c_{\theta_w}$ of the $Z$ fermion couplings.

%We next include corrections from the flavorons as well.
The flavoron models affect the low energy predictions through the
$Z$-fermion coupling shifts described in Section 2. The leading order
shifts as a function of the model parameters $\delta g_i(\kappa, M_F)$ are
easily calculable from the tree level expressions in (\ref{Vi}) for the
electroweak observables ($M_W$ is independent of the shifts to the Z-fermion 
couplings).  The models also give rise to shifts in the
$\Delta \rho$ ($T$) parameter.

% which we include through  (\ref{ST})

We include one non-$Z$-pole measurement in our fit, the lattice determination
of $\alpha_s$ \cite{pdg}
\beq
\alpha^{\rm lattice}_s(M_Z) =  0.117 \pm 0.003
\eeq
This constraint is important for our fit since the vertex corrections
in some of our models (in particular the universal coloron model) mimic a shift
in $\alpha_s(M_Z)$. Fitting to the lattice data ensures we do not 
produce an unacceptable value for $\Lambda_{QCD}$.
The value of $\alpha_s(M_Z)$ is, by these standards, imprecisely determined
and therefore we include possible variation in the input parameter 
$\alpha_s(M_Z)$ through $\delta \alpha_s$, a parameter which we will
include in the fit. We include the effect $\delta \alpha_s$ 
on electroweak measurements by modifying the
quark copulings\footnote{The
leading order QCD correction to the hadronic branching ratios
are $\Gamma_h = \Gamma^0_h (1 + \alpha_s/\pi)$.}

\beq 
\delta g_{q} = {e \over s_\theta c_\theta} 
(T_3 - Q s^2_\theta )\,{\delta \alpha_s \over  2\pi} 
\eeq

We perform the fit as follows. For a given model and choice of $M_F$ 
and $m_h$ we
have predictions for the shifts in the SM quantities in (\ref{Vi}) 
as a function of
$\kappa_F$ and $\delta \alpha_s$.  We calculate $\chi^2$ 
%
%\beq 
%\chi^2 = \sum_{EW variables, V_i} \left( {\rm pull}_i  - {\delta V_i 
%\over V_i^{SM}} {V^{\rm measured}_i \over \sigma_i} \right)^2
%\eeq
%
and place a limit on $\kappa_F$ by finding the $95\%$ confidence value
of $\chi^2$ allowing $\delta \alpha_s$ to take a value that minimizes
$\kappa_F$. The number of degrees of freedom is 14 - the number of fitted 
variables (16) minus 2 for the 2 fitted quantities. 

\section{Models and Limits}

We present five models as examples of flavoron physics and give their 
limits from the EW precision fit.

\subsection{Universal Coloron}

The universal coloron model \cite{unicol} 
at high energies has the gauge symmetry of the 
SM with regards the SM fermions but, in addition, an extra SU(3) gauge group
acting on a new fermion or scalar 
sector. That new sector also transforms under the proto-color group. 
It is assumed that dynamics
occurring in the new sector makes all the additional matter
massive
and induces a vev ($V$) for an effective higgs in the $(3, \bar{3})$ of the 
proto-color group and the additional SU(3).  
The two sets of SU(3) gauge bosons mix
through the mass matrix
\beq
(A^\mu, B^\mu) \left( \begin{array}{cc} g_{cp}^2 & -g_{cp} g_F 
\\ -g_{cp} g_F &g_F^2
\end{array} \right)V^2 \left( \begin{array}{c} A_\mu \\ B_\mu \end{array} \right)
\eeq
which is familiar and diagonalize to
\beq
(X^\mu, G^\mu) \left( \begin{array}{cc} g_{cp}^2 + g_F^2& 0 \\ 0 & 0
\end{array} \right)V^2 \left( \begin{array}{c} X_\mu \\ G_\mu \end{array} \right)
\eeq
where 
\beq
\left( \begin{array}{c} A^\mu \\ B^\mu \end{array} \right) =
\left( \begin{array}{cc} \cos \theta_F & -\sin \theta_F  \\ \sin \theta_F &
\cos \theta_F 
\end{array} \right) \left( \begin{array}{c} G_\mu \\ X_\mu \end{array} \right)
\eeq
with 
\beq 
\sin \theta_F = {g_{cp} \over \sqrt{g_{cp}^2 + g_F^2}}, \hspace{1cm}
\cos \theta_F = {g_F \over \sqrt{g_{cp}^2 + g_F^2}}
\eeq

Here the $G^\mu$ are the gauge bosons of the unbroken SU(3), which we
identify with QCD, and there is a massive
color octet of gauge bosons ($X^\mu$) that also couples to the SM quarks.
The low energy QCD coupling, with the standard generator normalization 
is given by
\beq 
g_c = {g_F g_{cp} \over \sqrt{(g_{cp}^2 + g_F^2)}}
\eeq
Note that this condition implies a minimum value for $\kappa_F
= g_F^2/4 \pi  \geq \alpha_s$ 
in order to obtain the physical color coupling. 
We take $\alpha_s(2\,{\rm TeV}) \simeq 0.09$.

The extra massive octet with mass $M_{F'}
= \sqrt{g_{cp}^2 + g_F^2} V$  couples as
\beq
g_{cp} A^{a\mu} \sum_f \bar{q}_f \gamma_\mu T^a q_f \rightarrow 
- g_c \cot \theta_F X^{a\mu} \sum_f \bar{q}_f \gamma_\mu T^a q_f 
+ ...
\eeq

If we assume that at low energies the massive gauge bosons may be approximated 
by a NJL model four fermion interaction
\beq
{\cal L}_{\rm eff} = - {g_c^2 \cot^2 \theta_F \over 2! M_{F'}^2}  
\left( \sum_f \bar{q}_f \gamma_\mu T^a q_f \right)^2
+ ...
\eeq
then the critical
coupling\footnote{Note that, defining the theory in terms of a
momentum-space cutoff $\Lambda$,  a four fermion interaction 
$G \bar{\psi}\psi \bar{\psi} \psi $ has a critical coupling $G_c = 2 \pi^2/\Lambda^2$
\cite{NJL}. The smaller value of $\cot\theta_{crit}$ given
in \cite{simmons,bertram} results from using the large-$N$ approximation and ignoring
the running in $\alpha_s$ from $M_Z$ to $M_{F'}={\cal O}(2\,{\rm TeV})$.}
for chiral symmetry breaking in that approximation is
\beq
\cot^2 \theta_{crit} = {2 N_c \pi \over (N_c^2-1)\alpha_s(2\,{\rm TeV})} = 26.2
\eeq

We may now calculate the corrections the model produces to the low energy
SM predictions. 
The universal coloron produces a universal shift to the $Z$ couplings
of all the quarks, from (\ref{vertex}), 
\beq
g_f \rightarrow g_f \left(1 +{\alpha_s \cot^2 \theta M_Z^2\over 6 \pi M_{F'}^{2}}
       {(N_c^2-1) \over 2 N_c} \ln \left({M_{F'}^2 \over M_Z^2} \right)\right)
\eeq
The correction to the coupling of the top quark 
would have to reflect the large top mass but 
the top does not enter into $Z$ pole observables. 
The contribution to the T parameter in the model is (\ref{tloop})
\beq
T = {N_c^2-1 \over 2} {3  m_t^4 \over 32 \pi^2 s_{\theta_w}^2 c_{\theta_w}^2
M_Z^2}  {\alpha_s \cot^2 \theta \over M_{F'}^{2}} 
\left[ \ln (M_{F'}^{2}/m_t^2) \right]^2
\eeq

The model has no $Z'$ to mix with the $Z$.  Performing a fit to the
$Z$-pole data as described in Section 3 produces the $95\%$ exclusion
curve shown in Fig 4. When the coupling is critical the massive octet
must lie above $3\,{\rm TeV}$ in mass.  In Fig 4 we also display the
precision bound reported in \cite{bertram}; that bound comes from the
weak $\rho$ parameter constraint $\Delta \rho \leq 0.4\%$ which assumes
the existence of new physics in the S parameter. As there are no large
contributions to the $S$ parameter, we derive a stronger bound. These
constraints should be compared to the best current experimental limits
from collider jet production \cite{bertram} also shown in Fig 4 - the
direct search limits are stronger than the precision measurement
constraint.

$\left. \right.$   \hspace{-0.2in}\ifig\prtbdiag{}
{\epsfxsize10truecm\epsfbox{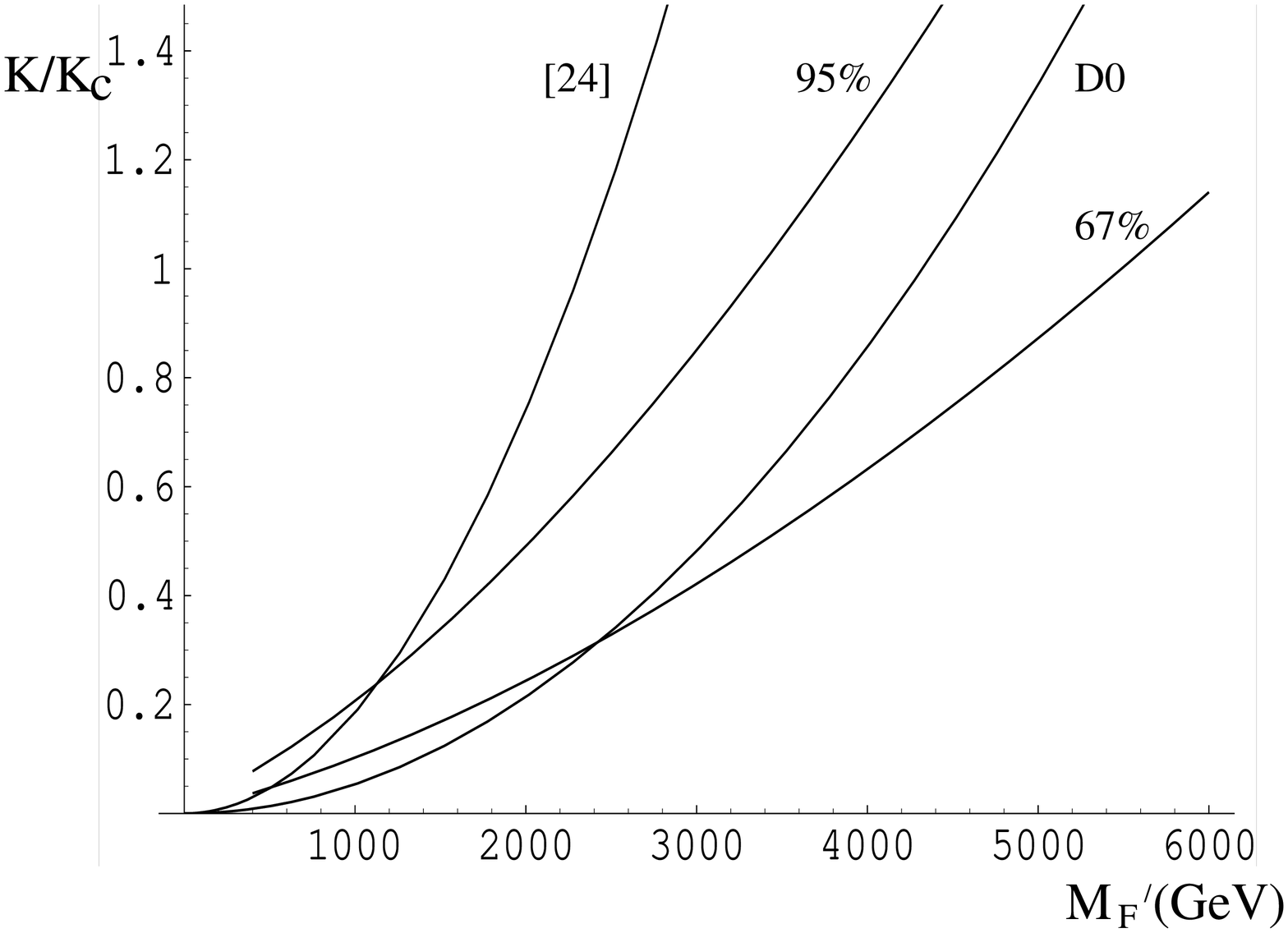}} \vspace{-1cm}
\begin{center}
  Figure 4: Exclusion curves in the $\kappa / \kappa_c$-$M_{F'}$ plane
  for the universal coloron model with a $100$ GeV higgs mass.  The
  curve marked [24] is the precision bound from ref. \cite{bertram}. The D0
  curve is the current $95\%$ confidence level
  experimental limit from collider jet production,
  also from \cite{bertram}. The remaining curves are our fit at the
  $95\%$ and $67\%$ confidence level.
\end{center}

\subsection{Chiral Top Color}

Top color \cite{tcol}  is very similar in the gauge sector to the universal
coloron model - above the flavor breaking scale there are two $SU(3)$
gauge groups that break to QCD and a massive color octet of gauge bosons. 
The SM fermions transform under the weaker (``proto-color group'') of
the two SU(3) groups except 
for the left handed top/bottom doublet, $\psi_L$, which 
transforms  under the stronger group. 
The top color group we consider is therefore chiral as in the 
see-saw model of \cite{tseesaw}.
The gauge-boson mixing is as for the universal coloron model.
Thus below the flavor breaking scale there are the
usual SM gluons plus one extra massive octet with mass $M_{F'}
= \sqrt{g_{cp}^2 + g_F^2} V$ and couplings
\beq 
g_{cp} A^{a \mu}    \bar{\psi}_L T^a \gamma_\mu \psi_L  + 
g_{tc} B^{a\mu} \sum_q \bar{q}  T^a \gamma_\mu q
\rightarrow 
-g_c \cot\theta_F  X^{a\mu} \bar{\psi}_L T^a \gamma_\mu \psi_L
+ g_c \tan \theta_F  X^{a\mu}  \sum_q \bar{q}  T^a \gamma_\mu q
+ ...
\eeq
where $q$ are any chiral  SM quark except  the left handed top/bottom 
multiplet. The critical coupling again
corresponds to $\cot^2 \theta = 26.2$.

The chiral top color model provides universal shifts to the $Z$ couplings
of all the quarks, excepting the top/bottom left handed doublet, of the form
\beq
g_q \rightarrow g_q \left(1 +{\alpha_s \tan^2 \theta M_Z^2\over 6 \pi M_F^{'2}}
       {(N_c^2-1) \over 2 N_c} \ln \left({M_{F'}^2 \over M_Z^2} \right)\right)
\eeq
The left handed bottom quark's $Z$ couplings receives the correction 
(\ref{vertex})
\beq
g_{b_L} \rightarrow g_{b_L} 
\left(1 +{\alpha_s \cot^2 \theta M_Z^2\over 6 \pi M_F^{'2}}
       {(N_c^2-1) \over 2 N_c} \ln \left({M_{F'}^2 \over M_Z^2} \right)\right)
\eeq
The contribution to the T parameter in the model is
\beq
T = {N_c^2-1 \over 2} { m_t^4 \over 32 \pi^2 s_{\theta_w}^2 c_{\theta_w}^2
M_Z^2} {\alpha_s (\cot^2 \theta + 2 \tan^2 \theta) \over M_F^{'2}} 
\left[ \ln(M_{F'}^{2}/m_t^2) \right]^2 
\eeq

The model has no $Z'$ that mixes with the $Z$. 
The result of performing the fit 
to the EW precision data is shown in Fig 5 and the $95\%$ confidence
limit is relatively low (1.3 TeV at critical coupling).

$\left. \right.$   \hspace{-0.2in}\ifig\prtbdiag{}
{\epsfxsize10truecm\epsfbox{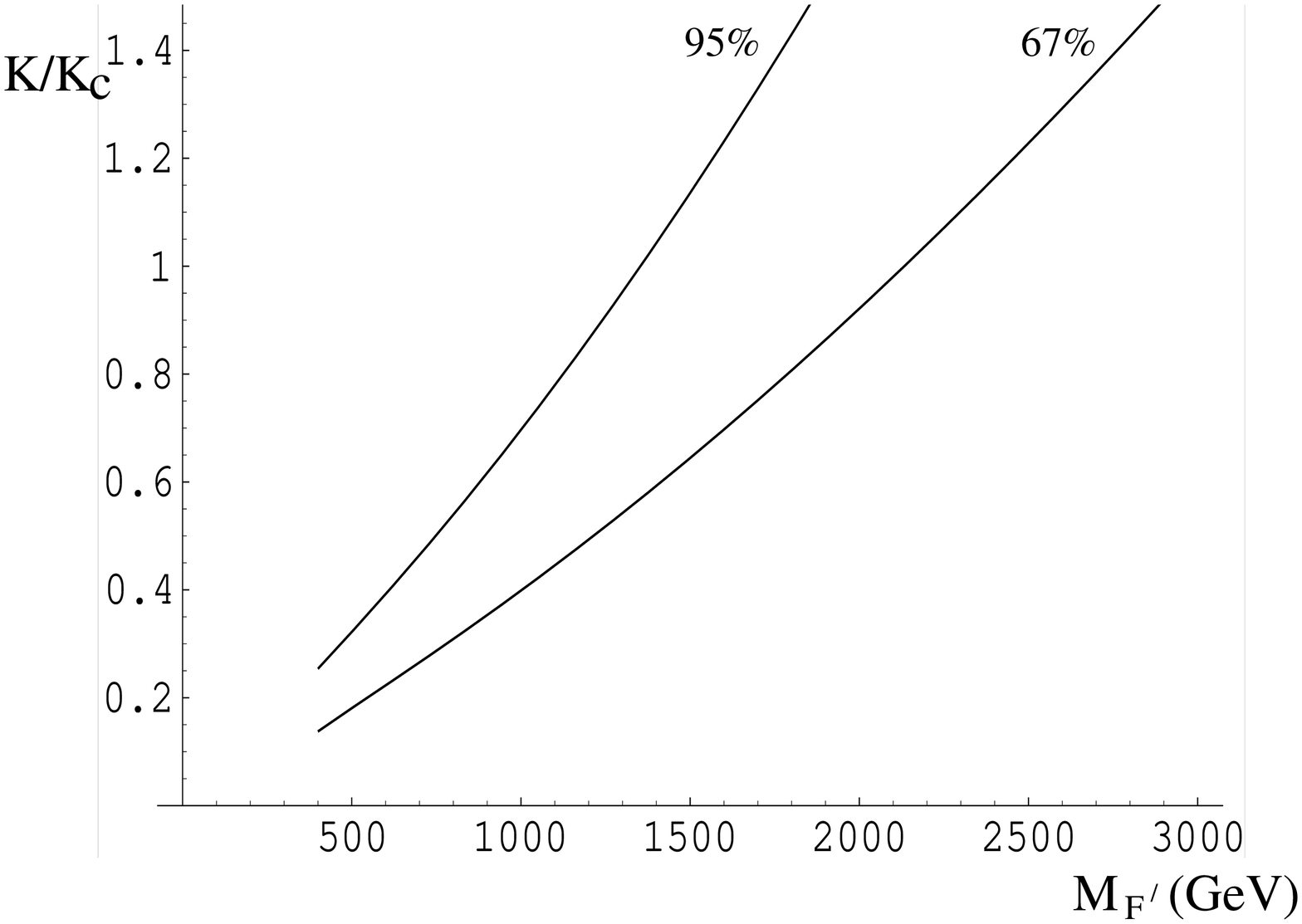}} \vspace{-1cm}
\begin{center}
Figure 5: The $95\%$ and $67\%$ exclusion curves in the 
$\kappa / \kappa_c$-$M_{F'}$ 
plane for the chiral top coloron model with a 100 GeV higgs mass.
\end{center}

\subsection{Chiral Quark Family Symmetry}

The gauging of the chiral family symmetry of the left handed quarks has
been motivated in technicolor \cite{georgi}, top condensate \cite{king}
and flavor universal see saw models \cite{fuseesaw}. The minimal
representative model has a gauged SU(3) family symmetry, in addition to
the SM interactions, acting on the three left handed quark doublets $Q =
((t,b)^i_L, (c,s)^i_L, (u,d)^i_L )$ where $i$ is a QCD index which
commutes with the family symmetry.  Note that gauging this symmetry
leaves the SM $U(3)^5$ global symmetry responsible for the GIM mechanism
\cite{ctsm} unbroken and the model is free of tree level FCNC
\cite{georgi}. The additional terms in the lagrangian are
\beq
{\cal L} = i g_F A^{\mu a} \bar{Q} \gamma_\mu T^a  Q 
\eeq
where $T^a$ are the generators of SU(3) corresponding to the three families. 
We assume that some massive sector
induces a vev for an effective higgs that completely breaks the SU(3) family
group giving the gauge bosons masses of order $M_F = g_F V$ where $V$ is the
mass scale associated with the symmetry breaking. There is no mixing with the
SM gauge group. 
 
If we assume that at low energies the massive gauge boson may be approximated 
by a NJL model with coupling $4 \pi \kappa / 2! M_{F}^2$ 
($\kappa = g_F^2/4\pi$),
then the critical
coupling for chiral symmetry breaking in that approximation is
\beq
\kappa_{crit} = {2 N \pi \over (N^2-1)} = 2.36
\eeq

The chiral quark familons give rise to universal corrections to all the 
left handed SM quark $Z$ couplings (\ref{vertex}). 
The only subtlety is including the top mass for isospin $+1/2$ quarks. 
\beq 
g_{+1/2_L} \rightarrow g_{+1/2_L} 
\left( 1 + {5 \kappa M_Z^2 \over 36 \pi M_F^2}
\ln \left({M_F^2 \over M_Z^2}\right) + {\kappa M_Z^2 \over 12 \pi M_F^2}
\ln \left({M_F^2 \over m_t^2}\right)\right)
\eeq
\beq
g_{-1/2_L} \rightarrow g_{-1/2_L} \left( 1 + {2 \kappa M_Z^2 \over 9 \pi M_F^2}
\ln \left({M_F^2 \over M_Z^2}\right) \right)
\eeq

The contribution to the T parameter in the model is (\ref{tloop})
\beq
T = {N_c \over 3} { m_t^4 \over 32 \pi^2 s_{\theta_w}^2 c_{\theta_w}^2
M_Z^2} {\kappa  \over M_F^2} \left[ 
\ln(M_F^2/m_t^2) \right]^2
\eeq

The result of the fit to EW data is shown in Fig 6. At critical coupling
the $95\%$ confidence level bound is approximately 2 TeV.
 
$\left. \right.$   \hspace{-0.2in}\ifig\prtbdiag{}
{\epsfxsize10truecm\epsfbox{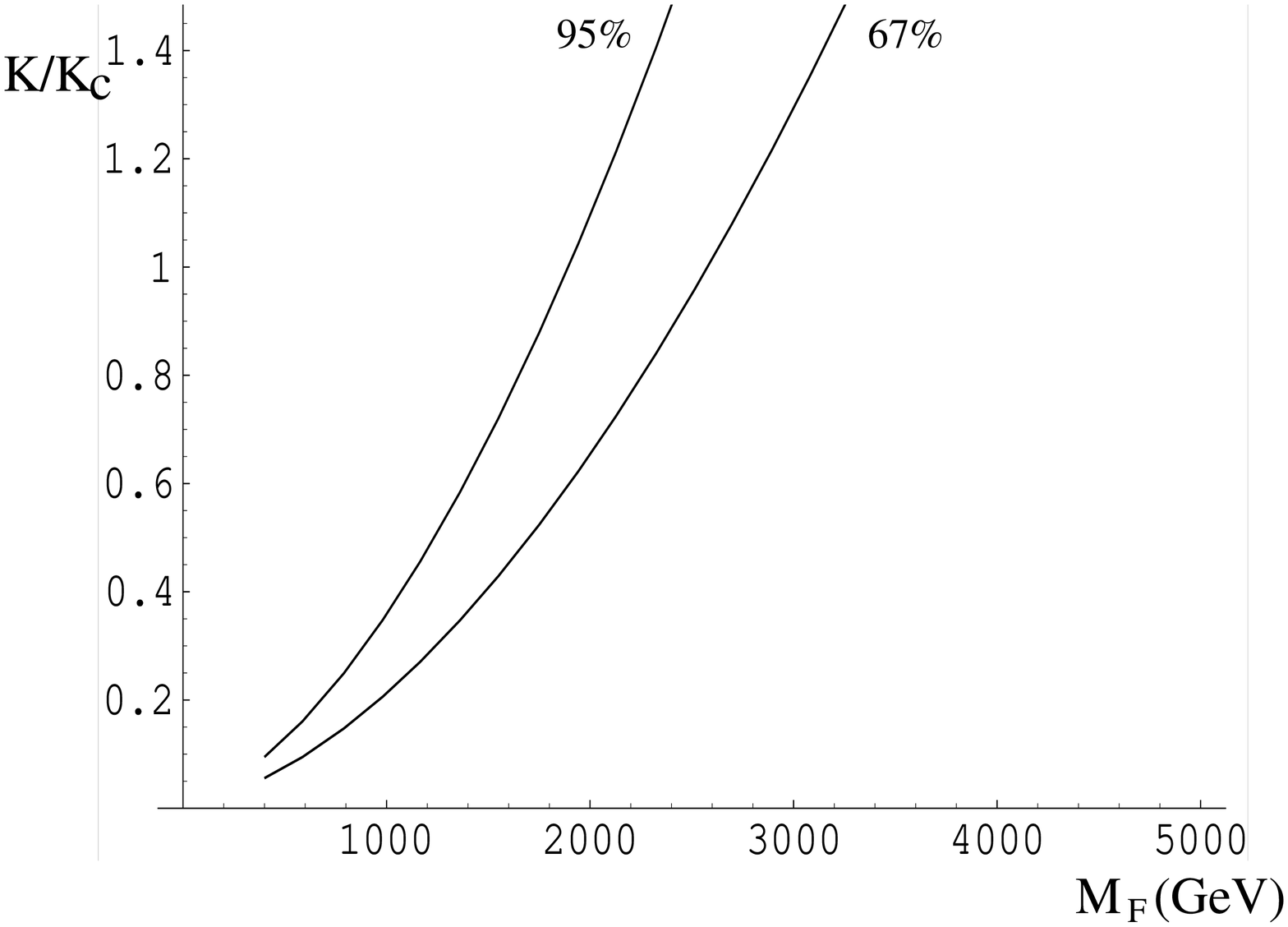}} \vspace{-1cm}
\begin{center}
Figure 6: The $95\%$ and $67\%$ exclusion curves in the 
$\kappa / \kappa_c$-$M_{F}$ 
plane for the gauged, chiral, quark family symmetry model with a 
100 GeV higgs mass.
\end{center}

\subsection{SU(9) Chiral Quark Flavor Symmetry}

We next consider a natural extension of the ideas of gauging the quark
family symmetry and chiral color symmetry. That is to gauge the full 
SU(9) symmetry of both the color and family multiplicity of the left handed
quarks.  Such a symmetry can be implemented as an extended technicolor 
gauge symmetry (in the spirit of \cite{randall}) or in quark universal 
seesaw models (as in \cite{fuseesaw}). Again the SM GIM is preserved
\cite{randall}
The SU(9) symmetry commutes with
the standard weak SU(2) gauge group and acts on the left handed quarks
\beq
Q_L = \left( (t,b)^r, (t,b)^b, (t,b)^g, (c,s)^r,... (u,d)^g \right)_L
\eeq
with $r,g,b$ the three QCD colors, as
\beq
{\cal L} = i g_F B^{a \mu} \bar{Q}_L \Lambda^a \gamma_\mu Q_L
\eeq
with $\Lambda^a$ the generators of SU(9) which include
\beq \label{su9gen}
{1 \over \sqrt{3}} 
\left( \begin{array}{ccc} 
              T^a & 0 & 0 \\ 0 & T^a & 0\\ 0 & 0 & T^a 
       \end{array}\right), 
{1 \over \sqrt{6}} 
\left( \begin{array}{ccc} 
              T^a & 0 & 0 \\ 0 & T^a & 0\\ 0 & 0 & -2 T^a 
       \end{array}\right),
{1 \over \sqrt{2}} 
\left( \begin{array}{ccc} 
              T^a & 0 & 0 \\ 0 & -T^a & 0\\ 0 & 0 & 0  
       \end{array}\right)
\eeq
where $T^a$ are the 8  3x3 QCD generators. SU(9) further contains
\beq 
{1 \over \sqrt{2}} 
\left( \begin{array}{ccc} 
              0 & T^a & 0 \\ T^a & 0 & 0\\ 0 & 0 & 0 
       \end{array}\right), 
{1 \over \sqrt{12}} 
\left( \begin{array}{ccc} 
              0 & 1 & 0 \\ 1 & 0 & 0\\ 0 & 0 & 0
       \end{array}\right),
{1 \over \sqrt{2}} 
\left( \begin{array}{ccc} 
             0 & -iT^a & 0 \\ iT^a & 0 & 0\\ 0 & 0 & 0  
       \end{array}\right),
{1 \over \sqrt{12}} 
\left( \begin{array}{ccc} 
             0 & -i & 0 \\ i & 0 & 0\\ 0 & 0 & 0  
       \end{array}\right)
\eeq
plus the two other similar sets mixing the remaining families. Finally there
are two diagonal generators
\beq
{1 \over \sqrt{12}} 
\left( \begin{array}{ccc} 
              1 & 0 & 0 \\ 0 & -1 & 0\\ 0 & 0 & 0 
       \end{array}\right), 
{1 \over \sqrt{36}} 
\left( \begin{array}{ccc} 
              1 & 0 & 0 \\ 0 & 1 & 0\\ 0 & 0 & -2
       \end{array}\right)
\eeq

The model must also contain the usual interactions of the right handed
quarks and leptons. These are included by a proto-color group that acts 
on the right handed quarks. 
We normalize the proto-color gauge bosons' couplings such that they have the 
same generators as the SU(9) bosons.
\beq
{\cal L} = {i\over \sqrt{3}} g_{cp} A^{\mu a} \bar{q}_R \gamma_\mu T^a q_R 
\eeq
The model must also include the usual SM hypercharge gauge boson.

At the flavor breaking scale we assume an effective higgs, 
that transforms as a 
$(9,\bar{9})$ under the $SU(9)_L \times SU(9)_R$ chiral flavor symmetry,
acquires a vev breaking the gauge symmetry to a single SU(3) that becomes
QCD. The majority of the SU(9) gauge bosons have mass $M_F=g_F V$. 
Eight of the SU(9) generators mix with the right
handed proto-color group. They correspond to the first generators in 
(\ref{su9gen}) 
which look like the usual QCD interactions of the left handed 
quarks.

The proto-gluons  and flavorons   mix through the mass matrix
\beq
(A^\mu, B^\mu) \left( \begin{array}{cc} g_{cp}^2 & -g_{cp} g_F 
\\ -g_{cp} g_F &g_F^2
\end{array} \right)V^2 \left( \begin{array}{c} A_\mu \\ B_\mu \end{array} \right)
\eeq
which diagonalizes to
\beq
(X^\mu, G^\mu) \left( \begin{array}{cc} g_{cp}^2 + g_F^2& 0 \\ 0 & 0
\end{array} \right)V^2 \left( \begin{array}{c} X_\mu \\ G_\mu \end{array} \right)
\eeq
where 
\beq
\left( \begin{array}{c} A^\mu \\ B^\mu \end{array} \right) =
\left( \begin{array}{cc} \cos \phi & -\sin \phi  \\ \sin \phi &
\cos \phi
\end{array} \right) \left( \begin{array}{c} G_\mu \\ X_\mu \end{array} \right)
\eeq
with
\beq 
\sin \phi = {g_{cp} \over \sqrt{g_{cp}^2 + g_F^2}}, \hspace{1cm}
\cos \phi = {g_F \over \sqrt{g_{cp}^2 + g_F^2}}
\eeq

The low energy QCD coupling, with the standard generator normalization 
is given by
\beq 
g_c = {g_F g_{cp} \over \sqrt{3(g_{cp}^2 + g_F^2)}}
\eeq
which implies that $\kappa_F \geq 3 \alpha_s(2\,{\rm TeV})$.

Thus the SM fermions' interactions with the massive color 
octet (with mass $M_{F'}
= \sqrt{g_{cp}^2 + g_F^2} V = M_F/c_\phi$) are given by
\beq
- g_c \tan \phi   X^{a \mu} \bar{q}_R \gamma_\mu T^a q_R +
g_c \cot \phi X^{a \mu} \bar{q}_L \gamma_\mu T^a q_L  
\eeq

If we assume that at low energies the massive gauge bosons may be
approximated by a NJL model with coupling $4 \pi \kappa / M_{F}^2$ (we
ignore the effects of the mixing of eight of the generators with
proto-color in this estimate which yields corrections of order
$g_{cp}/g_F$)) then the critical coupling for chiral symmetry breaking
in that approximation is
\beq
\kappa_{crit} = {2 N \pi \over (N^2-1)} = 0.71
\eeq

We may now calculate the deviations in low energy parameters in the 
SU(9) flavoron theory. The fermion $Z$ vertices are corrected to (\ref{vertex})
\begin{eqnarray}
\Delta g_{u_L} & = & g_{u_L} {\kappa M_Z^2\over 6 \pi M_F^2} \left[ \left(
         {5 N_c \over 6} + { c^4_{\phi} (N_c^2-1) \over 6 N_c}
          \right)\ln \left({M_F^2 \over M_Z^2} \right) 
          +{N_c \over 2} \ln \left({M_F^2 \over m_t^2} \right)\right]\\
&& \nonumber \\
\Delta g_{u_R} & = & g_{u_R}{\kappa M_Z^2\over 6 \pi M_F^2} \left[
           { s_\phi^4 (N_c^2-1) \over 6 N_c} \right]\ln \left({M_F^2 \over M_Z^2} \right)\\
&& \nonumber \\
\Delta g_{d_L} & = & g_{d_L}{\kappa M_Z^2 \over 6 \pi M_F^2} \left[
      {4 N_c \over 3} + { c_\phi^4 (N_c^2-1) \over 6 N_c }
          \right]\ln \left({M_F^2\over M_Z^2} \right)\\
\Delta g_{d_R} & = & g_{d_R}{\kappa M_Z^2\over 6 \pi M_F^2} \left[
           { s_\phi^4 (N_c^2-1) \over 6 N_c }\right]\ln \left({M_F^2 \over M_Z^2} \right)
\end{eqnarray}
where we have neglected the difference between $M_F$ and $M_{F'}$ 
in the logarithms.

The contribution to the T parameter in the model is (\ref{tloop})
\beq
T =
{ m_t^4 \over 32 \pi^2 s_{\theta_w}^2 c_{\theta_w}^2
M_Z^2}   N_c \left(  {\kappa \over M_F^2} + {4 \over 3} {\alpha_s (\cot^2 \phi
+ 2 \tan^2 \phi) 
\over M_{F}^2 } \right) \left[ 
\ln(M_F^2/m_t^2) \right]^2
\eeq

Again there is no mixing with the $Z$ boson.

The results of the fit to the electroweak data are displayed in Fig 7
and the $95\%$ confidence level
limit on the model at its critical coupling is a little below 2 TeV.

$\left. \right.$   \hspace{-0.2in}\ifig\prtbdiag{}
{\epsfxsize12truecm\epsfbox{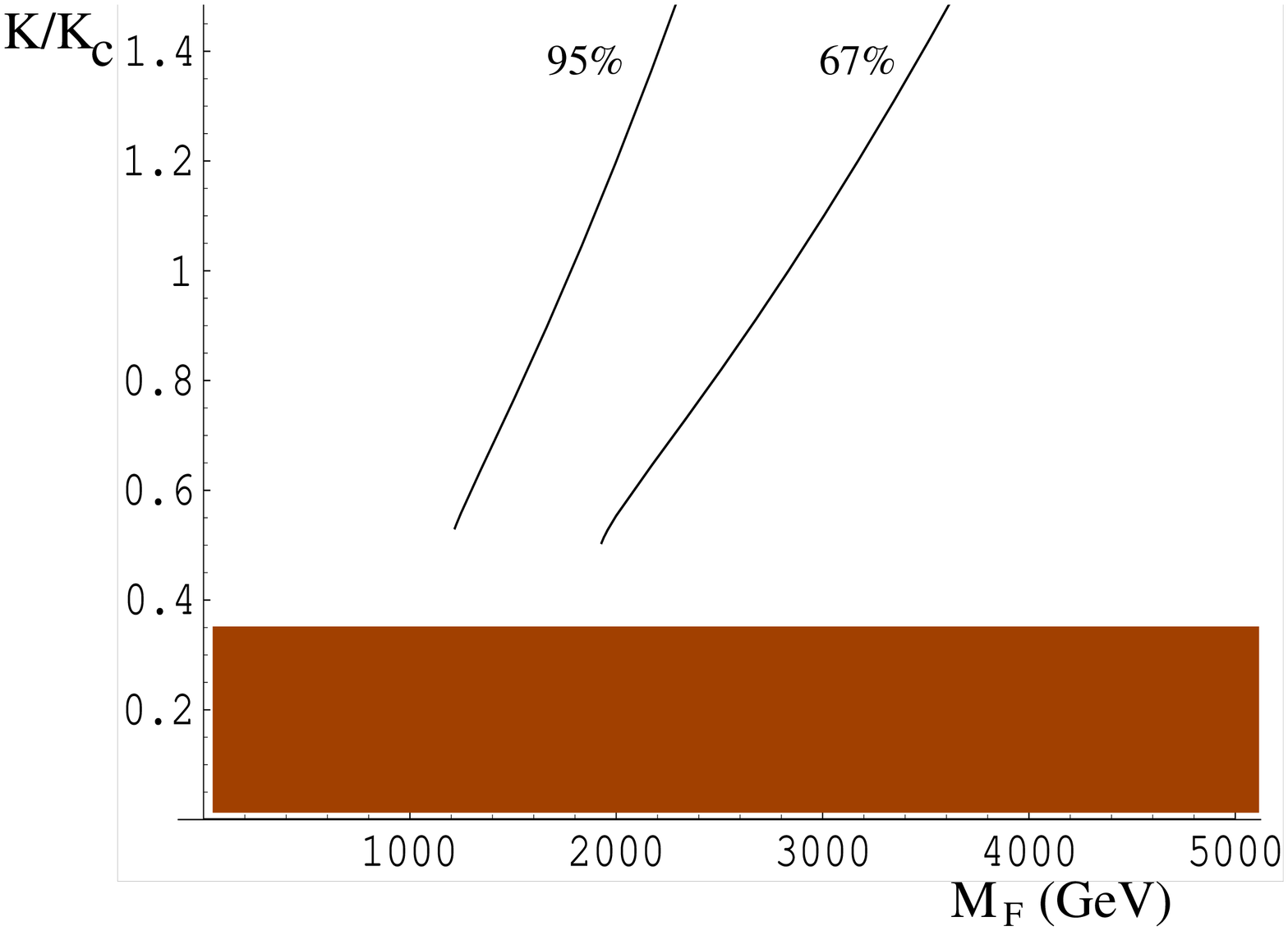}} \vspace{-1cm}
\begin{center}
Figure 7: The $95\%$ and $65\%$ exclusion curves in the 
$\kappa / \kappa_c$-$M_{F}$ 
plane for the gauged chiral quark SU(9) flavor symmetry model
with a 100 GeV higgs mass. The shaded
region is excluded by the requirement that the measured value of
$\alpha_s$ is obtained.
\end{center}

\subsection{SU(12) Chiral Flavor Symmetry}

The final model we consider is one in which we gauge the
full SU(12) flavor symmetry of all the left handed SM fermion
doublets \cite{randall,fuseesaw}
\beq
Q_L = \left( (t,b)^r, (t,b)^b, (t,b)^g, (\nu_\tau, \tau), 
(c,s)^r,... (\nu_e,e) 
\right)_L
\eeq
The flavor gauge interactions act as
\beq
{\cal L} = i g_F B^{a \mu} \bar{Q}_L \Lambda^a \gamma_\mu Q_L
\eeq
with $\Lambda^a$ the generators of SU(12), which may be conveniently  
broken down into the following groupings
\beq \label{su12gen}
{1 \over \sqrt{3}} 
\left( \begin{array}{ccc} 
              P^a & 0 & 0 \\ 0 & P^a & 0\\ 0 & 0 & P^a 
       \end{array}\right), 
{1 \over \sqrt{6}} 
\left( \begin{array}{ccc} 
              P^a & 0 & 0 \\ 0 & P^a & 0\\ 0 & 0 & -2 P^a 
       \end{array}\right),
{1 \over \sqrt{2}} 
\left( \begin{array}{ccc} 
              P^a & 0 & 0 \\ 0 & -P^a & 0\\ 0 & 0 & 0  
       \end{array}\right)
\eeq
where $P^a$ are the fifteen 4x4 Pati-Salam generators consisting of
eight 3x3 blocks that correspond to the QCD generators, six step
operators between the quarks and leptons and the diagonal generator
$1/\sqrt{24}$ diag$(1,1,1,-3)$. SU(12) further contains
\beq 
{1 \over \sqrt{2}} 
\left( \begin{array}{ccc} 
              0 & P^a & 0 \\ P^a & 0 & 0\\ 0 & 0 & 0 
       \end{array}\right), 
{1 \over \sqrt{16}} 
\left( \begin{array}{ccc} 
              0 & 1 & 0 \\ 1 & 0 & 0\\ 0 & 0 & 0
       \end{array}\right),
{1 \over \sqrt{2}} 
\left( \begin{array}{ccc} 
             0 & -iP^a & 0 \\ iP^a & 0 & 0\\ 0 & 0 & 0  
       \end{array}\right),
{1 \over \sqrt{16}}
\left( \begin{array}{ccc} 
             0 & -i & 0 \\ i & 0 & 0\\ 0 & 0 & 0  
       \end{array}\right)
\eeq
plus the two other similar sets mixing the remaining families. Finally there
are two diagonal generators
\beq
{1 \over \sqrt{16}} 
\left( \begin{array}{ccc} 
              1 & 0 & 0 \\ 0 & -1 & 0\\ 0 & 0 & 0 
       \end{array}\right), 
{1 \over \sqrt{48}} 
\left( \begin{array}{ccc} 
              1 & 0 & 0 \\ 0 & 1 & 0\\ 0 & 0 & -2
       \end{array}\right)
\eeq

To ensure the SM gauge groups emerge at low energies, we must again introduce
a proto-color group as in the SU(9) model above. The first 8 generators of
SU(12) in (\ref{su12gen}) are the same as those in the SU(9) model 
(\ref{su9gen}) and, hence, the discussion of the mixing between the proto-color
and the 8 SU(12) gauge bosons follows the discussion in the SU(9) model 
exactly.

In the SU(12) model, though, we must also include a proto-hypercolor gauge
boson. The Pati-Salam diagonal generator in the first set of generators in 
(\ref{su12gen}) is the traditional generator for the hypercharge boson's 
coupling to left handed fermions. The proto-hypercharge gauge boson
therefore only couples to the right handed fermions. We write the couplings
in each case in terms of the standard normalized hypercharges ($Q=T_3 + Y$)
and with the indicated normalizations of the couplings
\beq
{\cal L} = i g_{yp}{Y_R \over \sqrt{2}} \bar{f}_R A^\mu f_R + 
i g_F {Y_L \over \sqrt{2}}  \bar{f}_L B^\mu f_L 
\eeq
where $Y_L$ and $Y_R$ are the left and right handed 
hypercharges ($Y_L + Y_R = Y$).
At the flavor breaking scale the two gauge bosons mix through the mass matrix
\beq
(A^\mu, B^\mu) \left( \begin{array}{cc} g_{yp}^2 & -g_{yp} g_F 
\\ - g_{yp} g_F &g_F^2
\end{array} \right)V^2 \left( \begin{array}{c} 
A_\mu \\ B_\mu \end{array} \right)
\eeq
which diagonalizes to
\beq
(Z^{'\mu}, Y^\mu) \left( \begin{array}{cc} g_{yp}^2 + g_F^2& 0 \\ 0 & 0
\end{array} \right)V^2 \left( \begin{array}{c} Z'_\mu \\ Y_\mu \end{array} \right)
\eeq
where 
\beq
\left( \begin{array}{c} A^\mu \\ B^\mu \end{array} \right) =
\left( \begin{array}{cc} \cos \omega & -\sin \omega  \\ \sin \omega &
\cos \omega
\end{array} \right) \left( \begin{array}{c} Y_\mu \\ Z'_\mu \end{array} \right)
\eeq
with
\beq 
\sin \omega = {g_{yp} \over \sqrt{g_{yp}^2 + g_F^2}}, \hspace{1cm}
\cos \omega = {g_F \over \sqrt{g_{yp}^2 + g_F^2}}
\eeq
Here $Y^\mu$ is the ordinary hypercharge gauge boson.

The low energy hypercharge coupling, with the standard normalization of
hypercharges is given by
\beq 
g_Y = {g_F g_{yp} \over \sqrt{2 (g_{yp}^2 + g_F^2)}}
\eeq
and $\kappa_F \geq  \alpha_Y(2\,{\rm TeV})/2$ which is a lesser constraint than the 
constraint from obtaining the correct low energy QCD coupling discussed above.

The massive eigenstate has mass $\sqrt{g_F^2 + g_{yp}^2} V
= M_F / c_\omega$
and couples to the SM fermions as
\begin{eqnarray} \label{su12couple} \nonumber 
{\cal L } &  = & -i g_Y \tan \omega Y_R Z^{'\mu} \bar{f}_R \gamma_{\mu} f_R  + 
i g_Y \cot \omega Y_L Z^{'\mu} \bar{f}_L \gamma_{\mu} f_L  \\
\nonumber &&\\
& = & i {e \over c_\theta c_\omega s_\omega} (Y_L - s^2_\omega Y)
Z^{'\mu} \bar{f} \gamma_{\mu} f
\end{eqnarray}
where we have used the familiar SM result $g_Y = e/ c_\theta$.

If we assume that at low energies the massive gauge boson may be approximated 
by a NJL model with coupling $4 \pi \kappa / 2! M_{F}^2$ then the critical
coupling for chiral symmetry breaking in that approximation is
\beq
\kappa_{crit} = {2 N \pi \over (N^2-1)} = 0.53
\eeq

Note that combined with the lower bound from the ability to reproduce 
the QCD coupling ($\kappa_F \geq 3 \alpha_s(2\,{\rm TeV}) \simeq 0.3$) 
there is a relatively small window of allowed couplings.

The chiral SU(12) flavoron gives rise to more complicated 
expressions for the shifts to
the $Z$ couplings because it mixes quarks and leptons. The shifts 
generated by flavoron exchange across the vertices, given by (\ref{vertex}),
are
\begin{eqnarray}
\Delta g_{u_L} & = & g_{u_L} {\kappa M_Z^2\over 6 \pi M_F^2} \left[ \left(
         {5 \over 2} + { c^4_{\phi} (N_c^2-1) \over 6 N_c}
          + { c_\omega^4 Y_{u_L}^2 \over 8}
          \right)\ln \left({M_F^2 \over M_Z^2} \right) \right.\\
&& \nonumber \left. +{N_c \over 2} \ln \left({M_F^2 \over m_t^2} \right)\right]
          + g_{\nu_L} {\kappa M_Z^2\over 6 \pi M_F^2} \left[
               {3 \over 2} \ln \left({M_F^2 \over M_Z^2} \right) \right]\\
&& \nonumber \\
\Delta g_{u_R} & = & g_{u_R}{\kappa M_Z^2\over 6 \pi M_F^2} \left[
           { s_\phi^4 (N_c^2-1) \over 6 N_c}
          + { s^4_\omega Y_{u_R}^2 \over 8} \right]\ln \left({M_F^2 \over M_Z^2} \right)\\
&& \nonumber \\
\Delta g_{d_L} & = & g_{d_L}{\kappa M_Z^2 \over 6 \pi M_F^2} \left[
      4   + { c_\phi^4 (N_c^2-1) \over 6 N_c }
          + { c_\omega^4 Y_{d_L}^2 \over 8 }
          \right]\ln \left({M_F^2\over M_Z^2} \right)\\  
&&          + g_{e_L} {\kappa M_Z^2\over 6 \pi M_F^2} \left[
                     {3 \over 2} \ln \left(M_F^2/M_Z^2 \right) \right]\\
\Delta g_{d_R} & = & g_{d_R}{\kappa M_Z^2\over 6 \pi M_F^2} \left[
           { s_\phi^4 (N_c^2-1) \over 6 N_c }
          + { s_\omega^4 Y_{d_R}^2 \over 8}  \right]\ln \left({M_F^2 \over M_Z^2} \right)\\
 && \nonumber      \\
\Delta g_{\nu_L} & = & g_{\nu_L}{\kappa M_Z^2 \over 6 \pi M_F^2} \left[     
    {4 \over 3} + { c^4_\omega Y_{\nu_L}^2 \over 8 }
         \right]\ln \left({M_F^2 \over M_Z^2} \right) 
         + g_{u_L}  {N_C \kappa M_Z^2 \over 6 \pi M_F^2}  
          \ln \left({M_F^2 \over M_Z^2} \right) \\
&& + 
          g_{u_L}  {N_C \kappa M_Z^2 \over 12 \pi M_F^2}  
          \ln \left({M_F^2 \over m_t^2} \right) \\
&& \nonumber \\
\Delta g_{e_L} & = & g_{e_L}{\kappa M_Z^2 \over 6 \pi M_F^2} \left[
      {4 \over 3} + { c^4_\omega Y_{e_L}^2 \over 8}
         \right]\ln \left({M_F^2 \over M_Z^2} \right) 
         + g_{d_L}  {3 N_C \kappa M_Z^2 \over 12 \pi M_F^2}  
          \ln \left({M_F^2 \over M_Z^2} \right) \\
&& \nonumber \\
\Delta g_{e_R} & = & g_{e_R}{\kappa M_Z^2 \over 6 \pi M_F^2} \left[
 { s^4_\omega Y_{e_R}^2 \over 8}
         \right]\ln \left({M_F^2 \over M_Z^2}\right)
\end{eqnarray}

The contribution to the T parameter from the diagram in Fig 2 
in the model is, from (\ref{tloop})
\beq
T = 
{ m_t^4 \over 32 \pi^2 s_{\theta_w}^2 c_{\theta_w}^2
M_Z^2}   N_c \left(   {\kappa \over M_F^2} 
+ {4 \over 3} {\alpha_s (\cot^2 \phi + 2 \tan^2 \phi) 
\over M_{F}^2 }  + {1 \over 9} {\alpha_Y (\cot^2 \omega + 2 \tan^2 \omega)
\over M_{F}^2 } \right) \left[ 
\ln(M_F^2/m_t^2) \right]^2
\eeq

Finally we must also include the effects from the mixing between the
proto-hypercolor group and the diagonal Pati-Salam generator. We must make
some assumption about the structure of the EW breaking sector; we assume that
EWS is broken by the equal vevs of a set of fields satisfying $\sum Y_L = 0$ as
would be the case if condensates of all the SM fermions were responsible for
the breaking (as in the model of \cite{fuseesaw}). Writing the standard model
$Z$ mass as
\beq
M_Z^2 = {e^2 \over s_\theta^2 c_\theta^2} \langle T_3 T_3 \rangle
\eeq
where $\langle T_3 T_3 \rangle$ is the expectation value of the EWS breaking
operator coupling to the $T_3$ currents, the mass mixing between the $Z$ and
the $Z'$ gauge bosons is given by
\beq
\Delta M^2 = {e^2 \over s_\theta c_\theta^2 s_\omega c_\omega} s_\omega^2  
\langle T_3 T_3 \rangle = {s_\theta s_\omega \over c_\omega} M_Z^2
\eeq
where we have used the coupling in (\ref{su12couple}), 
the assumption $\sum Y_L =0$,
and $\langle Q T_3 \rangle =0$. The resulting shifts in $Z$ couplings
are given by
\beq
\delta g_f = - {e \over s_\theta c_\theta}{ M_Z^2 \over M_F^2} s_\theta^2
(Y_L - s^2_\omega Y)
\eeq
and the contribution to T is
\beq
T = {s_\theta^2 s_\omega^2 \over \alpha_{EW} } {M_Z^2\over M_F^2}
\eeq

The result of the fit to the EW data is displayed in Fig 8 and the 
$95\%$ confidence level limit at
critical coupling is $2\,{\rm TeV}$.

$\left. \right.$   \hspace{-0.2in}\ifig\prtbdiag{}
{\epsfxsize12truecm\epsfbox{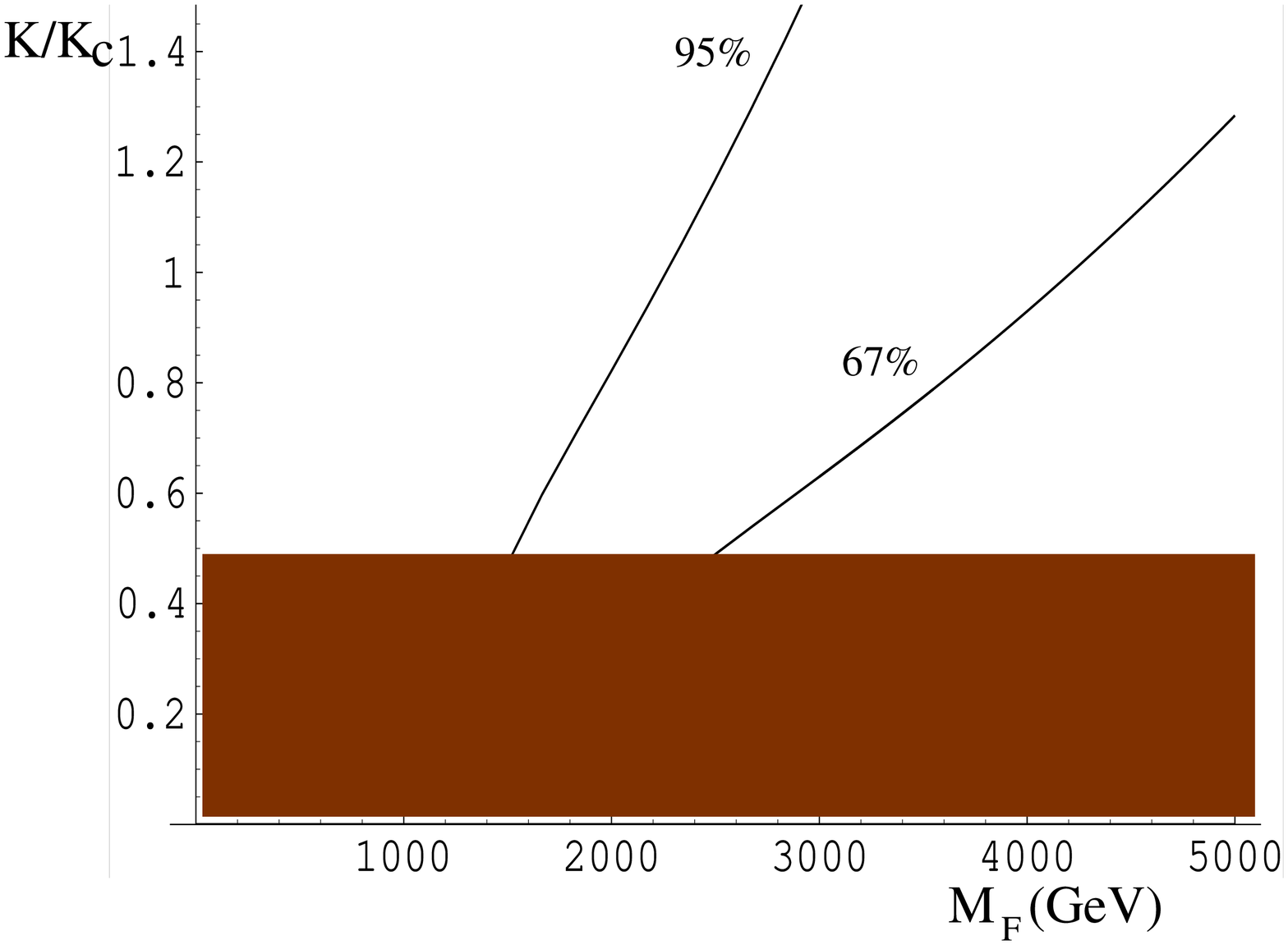}} \vspace{-1cm}
\begin{center}
Figure 8:
The $95\%$ and $67\%$ exclusion curves in the $\kappa / \kappa_c$-$M_{F}$ 
plane for the gauged chiral SU(12) flavor symmetry model with a 100 GeV
higgs mass. The shaded
region is excluded by the requirement that the measured value of
$\alpha_s$ is obtained.
\end{center}

\section{Robustness of Bounds}

In this section we address the robustness of the bounds to a number of 
factors. We will discuss how the fit changes as the higgs mass varies
and  how strongly the fits depend on the bottom quark asymmetries
and left right asymmetry measurement from SLD.
Finally the possible influence on the bounds from the presence of
additional new physics is investigated using the example of the SM
fermions mixing with heavy EW singlet fermions.

\subsection{Variation in the Higgs Mass}

The fits so far performed have assumed a higgs with a mass of $100$ GeV.
It is interesting to study the robustness of the limits to changes in
the higgs mass. In fact in many of the dynamical models of EWS breaking
we have used to motivate the existence of flavorons, the higgs is
expected to be somewhat heavier; typically in flavor universal see-saw
models \cite{fuseesaw} the higgs mass is expected to be of order 400
GeV. In the SM such a mass would be excluded but with the additional
flavoron physics it is potentially allowed. It is worth noting that the 
230 GeV upper bound on the higgs mass in the standard model is largely 
the result of the SLD measurement of the left right asymmetry. Removing that
piece of data raises the limit to of order 400 GeV. The influence of the
SLD data on the flavoron bounds, however, is small. This can be seen 
in Fig 9 where the limit on the universal coloron model is shown if the 
pull of the data from the fit was zero.

$\left. \right.$   \hspace{-0.2in}\ifig\prtbdiag{}
{\epsfxsize12truecm\epsfbox{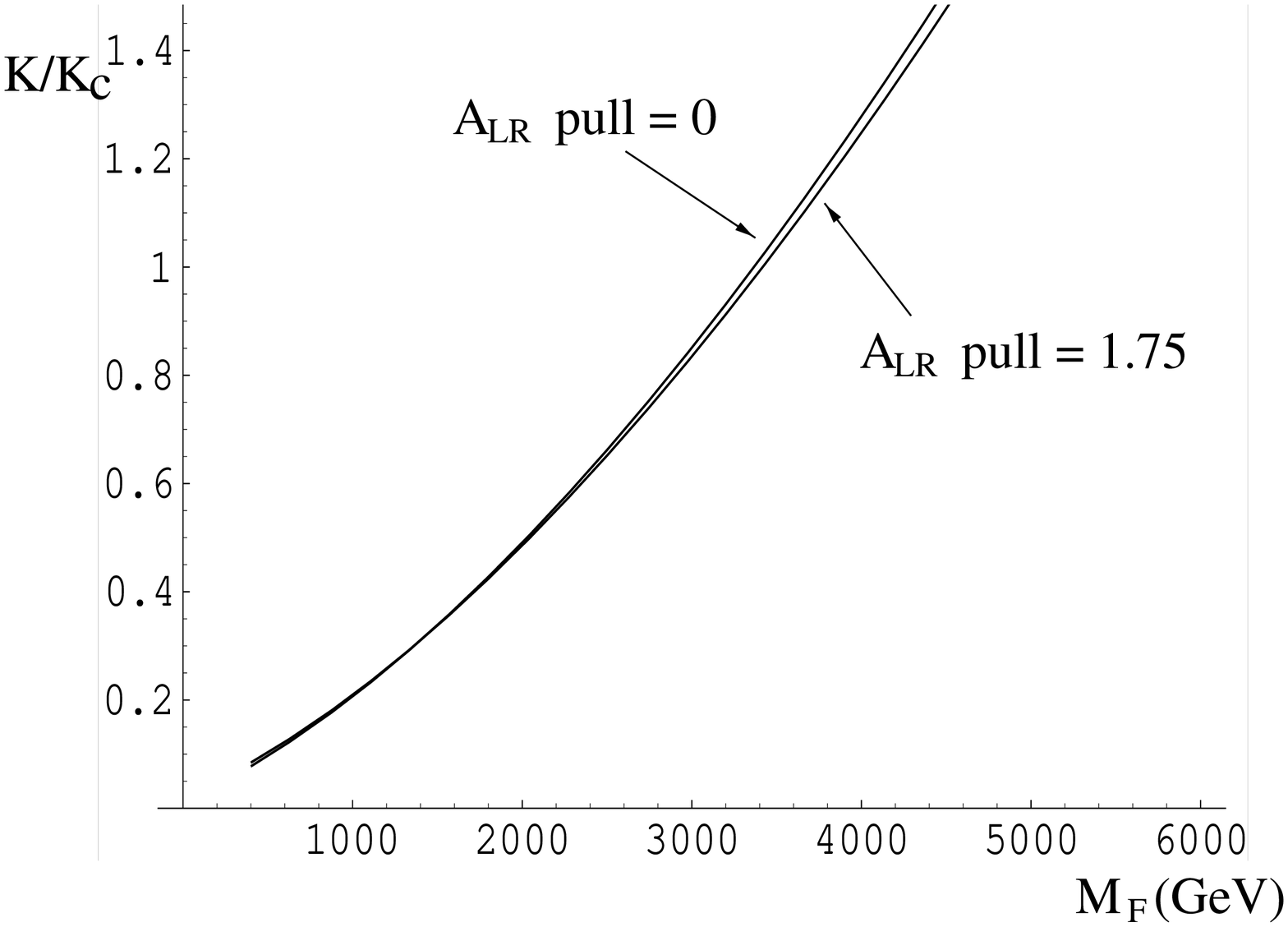}} \vspace{-1cm}
\begin{center}
Figure 9: The $95\%$ exclusion curves in the $\kappa / \kappa_c$-$M_{F}$ 
plane for the universal coloron model with a 100 GeV higgs mass and
varying $A_{LR}$ pull. 
\end{center}

As a first example of the higgs mass dependence, 
we show in Fig 10 the $95\%$ confidence level bounds
on the chiral quark familon model with varying higgs mass. The bound
on the flavoron mass scale at critical coupling 
is relatively insensitive to such changes. 
The model has little effect on the higgs mass 
upper bound, which in our fit is  230 GeV. The higgs mass dependence 
of the SU(9) and SU(12) chiral flavor symmetry models follow
this same pattern as the familon model.

$\left. \right.$   \hspace{-0.2in}\ifig\prtbdiag{}
{\epsfxsize12truecm\epsfbox{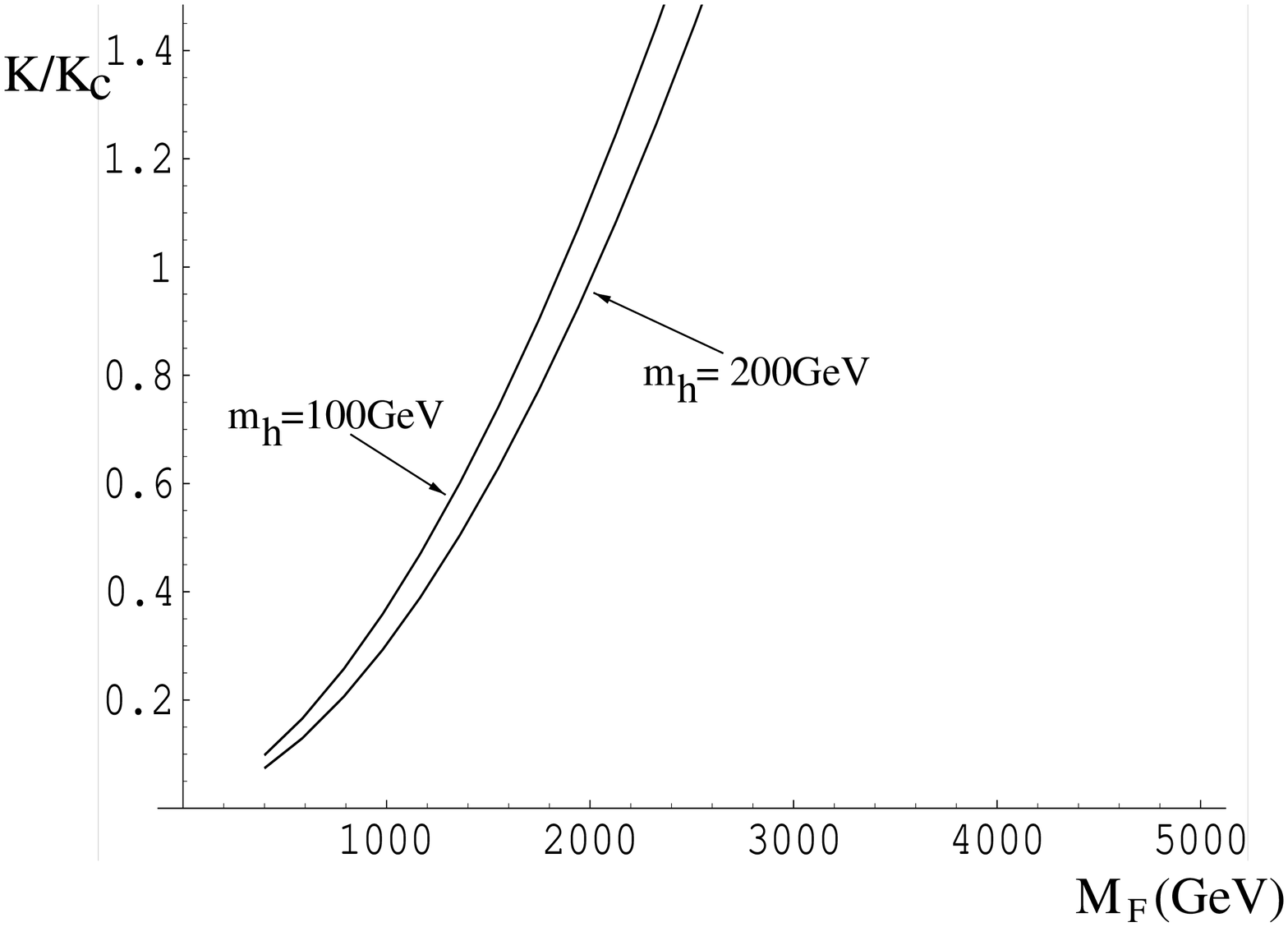}} \vspace{-1cm}
\begin{center}
Figure 10: The $95\%$ exclusion curves in the $\kappa / \kappa_c$-$M_{F}$ 
plane for the chiral quark family symmetry model with varying higgs mass. 
\end{center}

The universal (and chiral top) coloron models have larger contributions
to the T parameter which lead to these models' preferring a heavy higgs
(which gives negative contributions to T).  The dependence on the higgs
mass is shown in Fig 11 for the universal coloron model. If the higgs is
light then the limits on the colorons rise. For higgs masses above 200
GeV the limits are relatively insensitive. Note that it is possible in
these models for the higgs mass to lie above the SM $95\%$ exclusion
limit - indeed, the upper higgs bound can be increased to 650 GeV (or well
over 1 TeV if the SLD data point is removed from the fit).
Larger higgs masses are {\it only} allowed in the presence of extra
gauge bosons and there is also a {\it lower bound} on the coupling of
the flavoron at these high higgs mass values. As an example in Fig 12 we
show the $95\%$ exclusion curve for a 400 GeV higgs.

$\left. \right.$   \hspace{-0.2in}\ifig\prtbdiag{}
{\epsfxsize12truecm\epsfbox{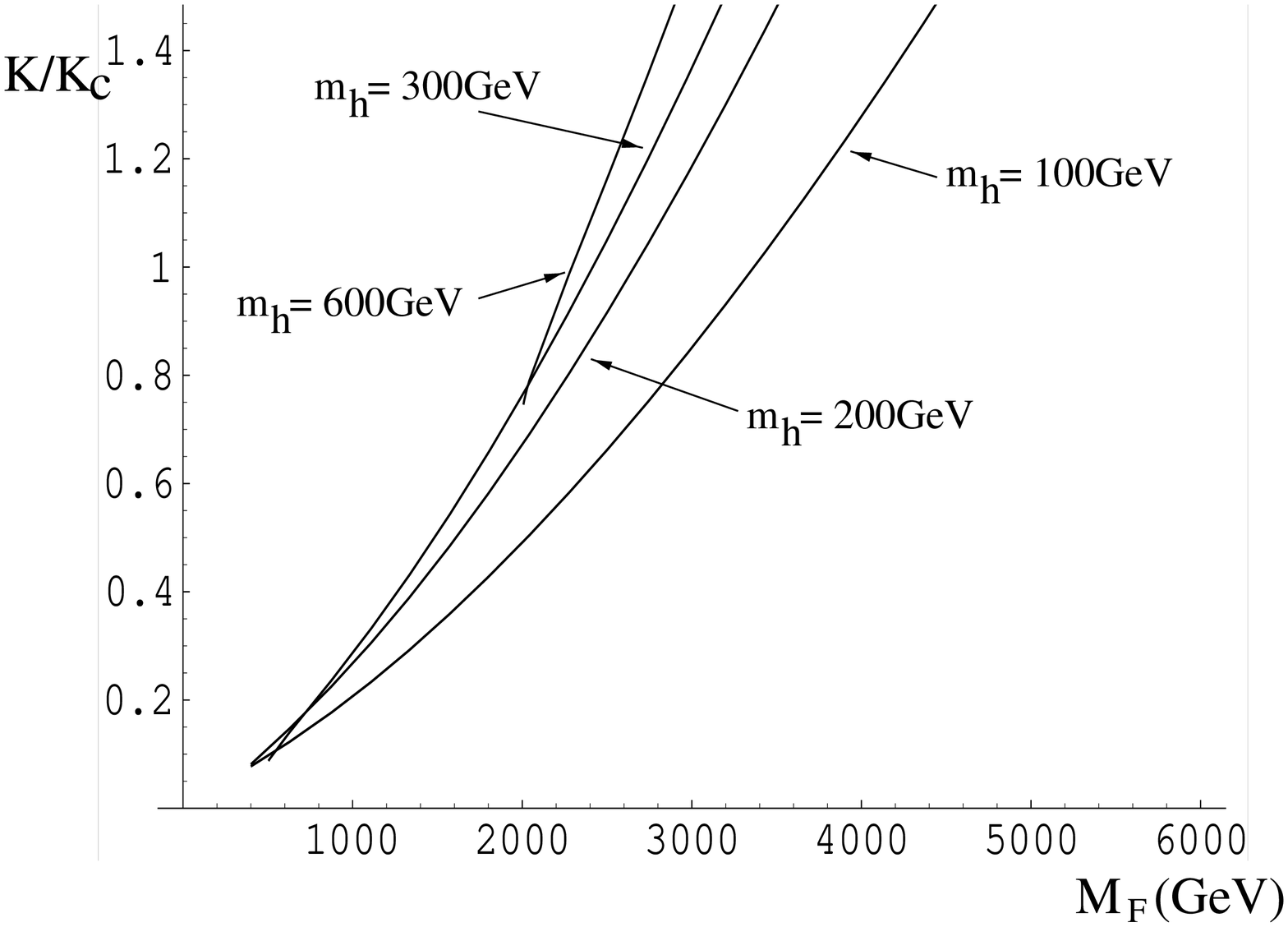}} \vspace{-1cm}
\begin{center}
  Figure 11: The $95\%$ exclusion curves from fits to Z-pole data in the
  $\kappa / \kappa_c$-$M_{F}$ plane for the universal coloron model with
  varying higgs mass.
\end{center}

$\left. \right.$   \hspace{-0.2in}\ifig\prtbdiag{}
{\epsfxsize12truecm\epsfbox{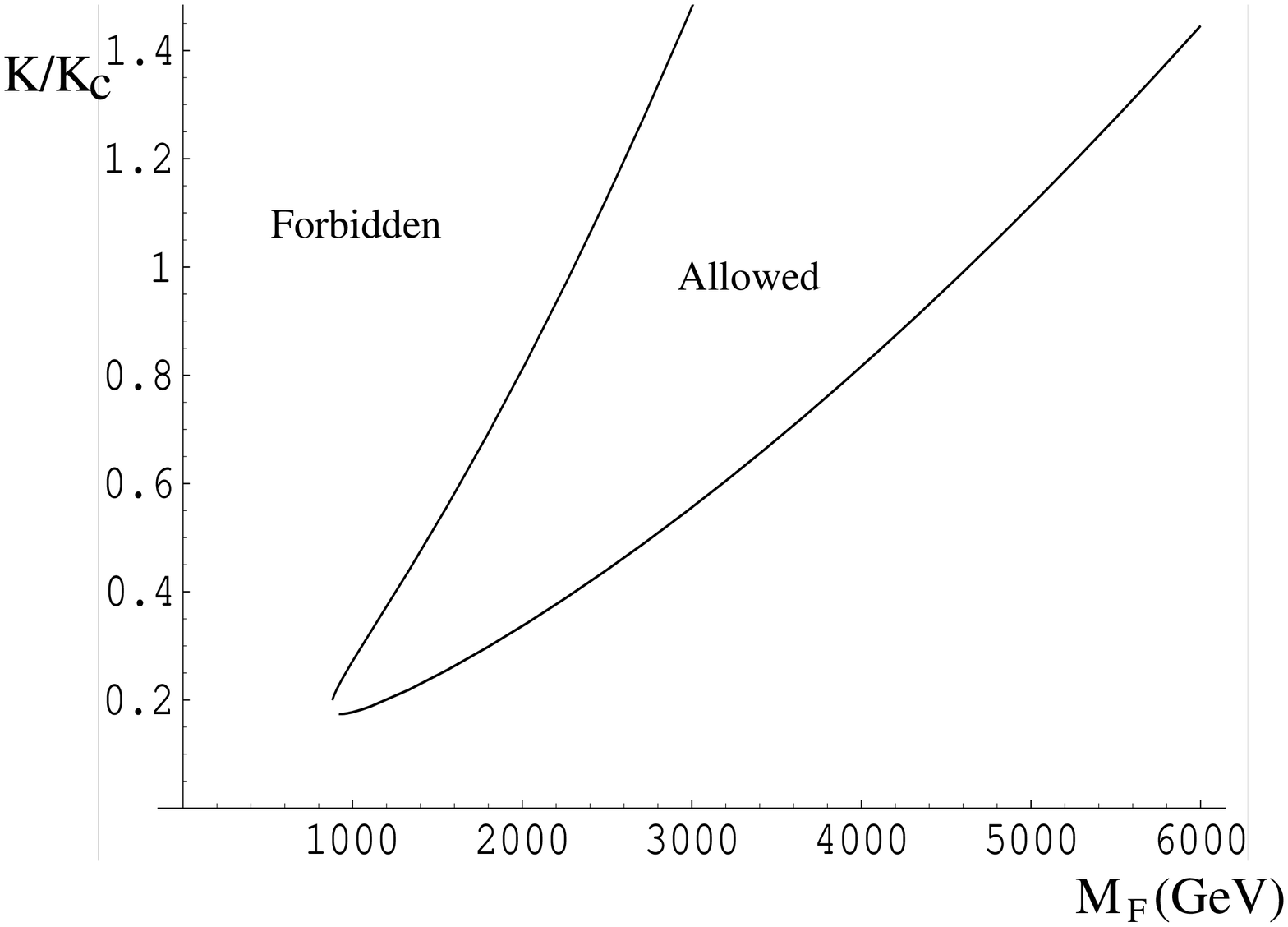}} \vspace{-1cm}
\begin{center}
  Figure 12: The $95\%$ exclusion curves in the $\kappa /
  \kappa_c$-$M_{F}$ plane for the universal coloron model with a 400 GeV
  higgs mass. Note the appearance of a {\it lower} bound on the coloron
  mass: a 400 GeV higgs is only allowed in the presence of interactions
  beyond the standard model.
\end{center}

\subsection{The Role of $A_b$ and $A_{fb}^{0,b}$}

A cursory overview of the data in (\ref{Vi}) reveals that the quantities
$A_b$ and $A_{fb}^{0,b}$ are those furthest from the SM predictions.
What role are these discrepancies playing in the fit? Although  the
experimental deviation is in the opposite direction to the shifts in these
quantities induced by the vertex corrections in (1), their role in
constraining the models is not too great. As an indication 
of the importance of these out lying measurements we show in Fig 13
what the bounds would have been in the universal coloron model
if the pulls in the SM fit had been zero. Relaxing these 
measurements weakens the bound although by only a few hundred GeV. The 
other flavoron models behave very similarly in this respect to the 
universal coloron  model. 

$\left. \right.$   \hspace{-0.2in}\ifig\prtbdiag{}
{\epsfxsize12truecm\epsfbox{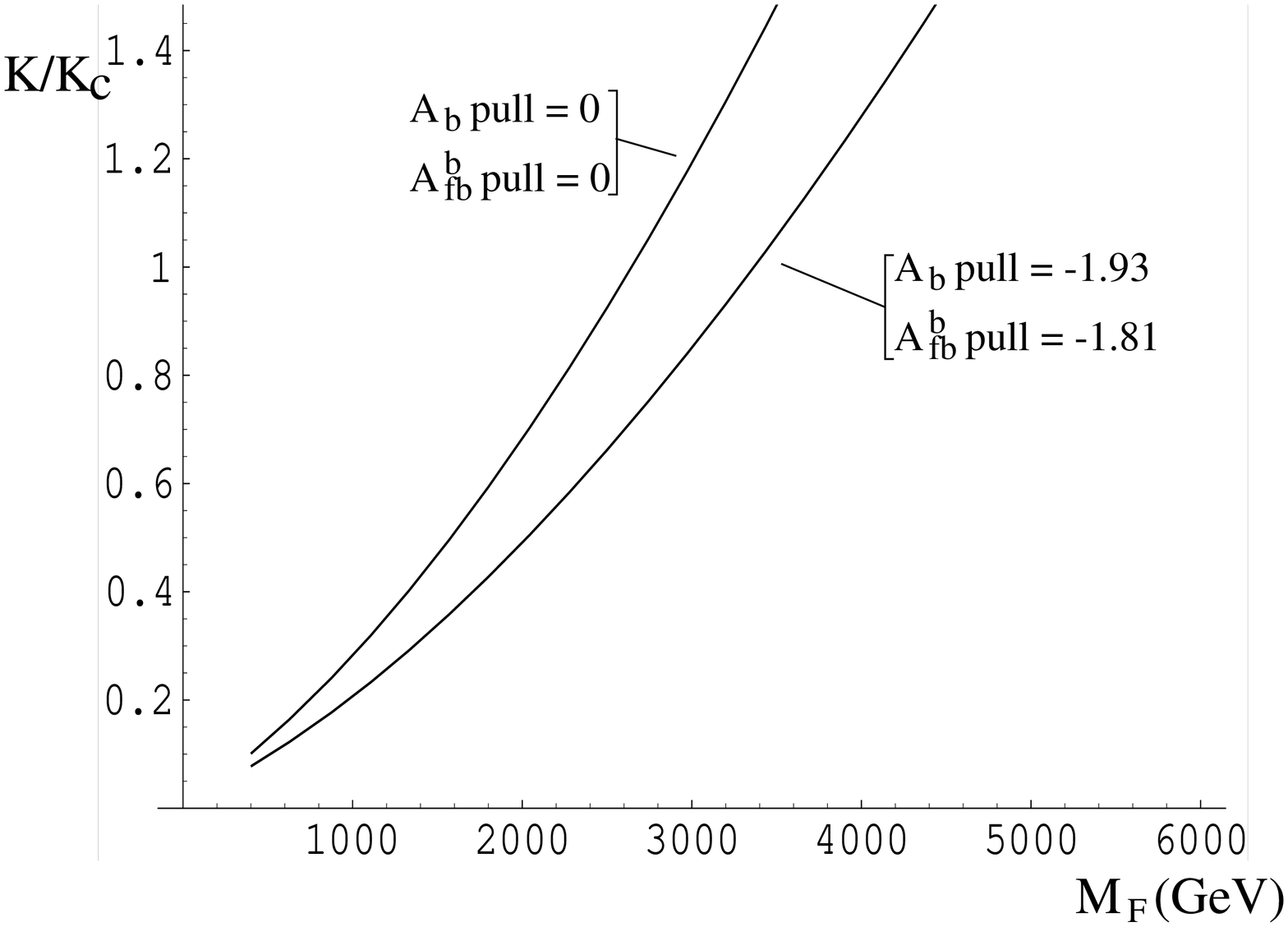}} \vspace{-1cm}
\begin{center}
Figure 13: The $95\%$ exclusion curves in the $\kappa / \kappa_c$-$M_{F'}$ 
plane for the universal coloron  model with a 100 GeV higgs mass
and varying $A_b$ and $A_{fb}^b$ experimental pull.  
\end{center}

\subsection{Flavorons and Fermion Mixing}

As an example of the role that extra new physics can play in changing the
bounds on flavorons, in this section we include the possibility of the 
SM fermions mixing
with massive electroweak singlet states. Fermion mixing is an integral part
of the top \cite{tseesaw} and flavor universal \cite{fuseesaw}
see-saw models. In these models EW symmetry is
broken by a condensate between the left handed SM fermions and massive, 
singlet Dirac fermions with the quantum numbers of 
the right handed SM fermions.  
Although these models have 
a decoupling limit, in which the mass of the EW singlet fermions
may be taken arbitrarily high, for the dynamics that 
breaks EW symmetry to involve the singlets we require that they are not more 
massive than the flavor gauge bosons.  

$\left. \right.$   \hspace{-0.2in}\ifig\prtbdiag{}
{\epsfxsize6truecm\epsfbox{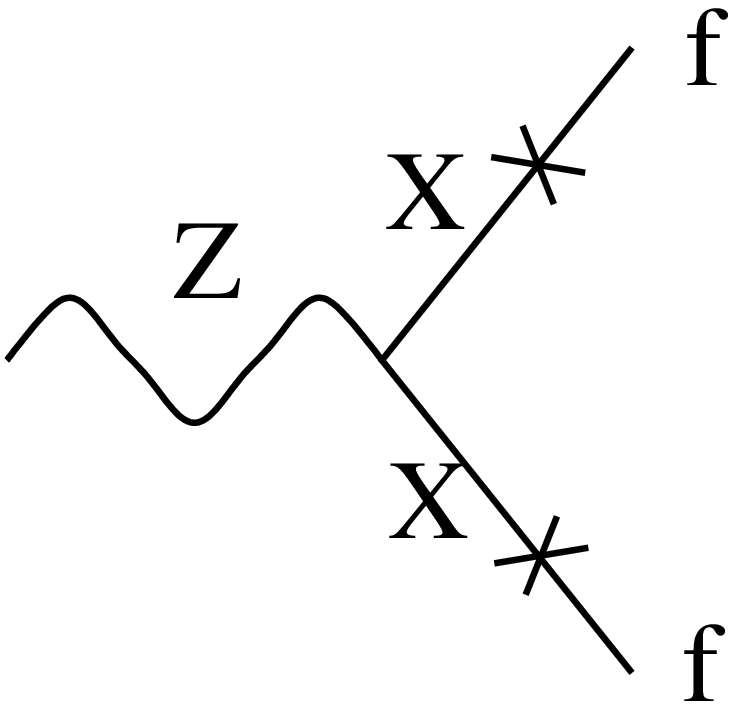}} \vspace{-1cm}
\begin{center}
Figure 14: The diagram correcting fermion (f) $Z$ vertices when the SM
fermions mix with massive singlets, $\chi$.
\end{center}

To include these effects we assume that all the SM fermions mix with
singlets (one pair for each type of massive standard model fermion) with
masses $M_F$. We assume that the mass mixing term is the same for all
the SM fermions and display the results of a fit to the EW data for mass
mixings, $M_{mix}$, between 50 GeV and 150 GeV. We treat the mixing in
perturbation theory through the graph in Fig 14 so the shifts to the SM
fermion $Z$ vertices are
\beq \label{mix}
\delta g_{f} \simeq -{e \over s_\theta c_\theta}
{Q_f s_{\theta_w}^2 } \left({M_{mix} \over M_F}\right)^2~,
\eeq
since $M_F \gg M_{mix}$.

In Fig 15 we display the results of the fit to the precision
data\footnote{Top-quark mixing with a singlet fermion also effects the
  one-loop contribution(s) to $\Delta\rho/T$ \cite{tseesaw}. This effect
  can be made relatively small, depending on the details of the model.
  We include here only the added, and potentially much larger, effect of
  fermion mixing. } for the chiral, quark familon symmetry model. The
fermion mixing gives rise to a sharp cut off in $M_F$ where the shifts
in the vertices from the mixing alone saturate the experimental bounds.
Above that scale the constraints on the flavorons are in fact reduced
slightly because the mixing vertex corrections (\ref{mix}) have the
opposite sign to those of the flavoron corrections (\ref{vertex}). As
the mixing mass rises to 200 GeV the bound from the mixing dominates.
The behavior of the familon model is indicative of the effect in the
other models where the effect is also small upto the scale of mixing
where that new physics dominates the bound.

$\left. \right.$   \hspace{-0.2in}\ifig\prtbdiag{}
{\epsfxsize12truecm\epsfbox{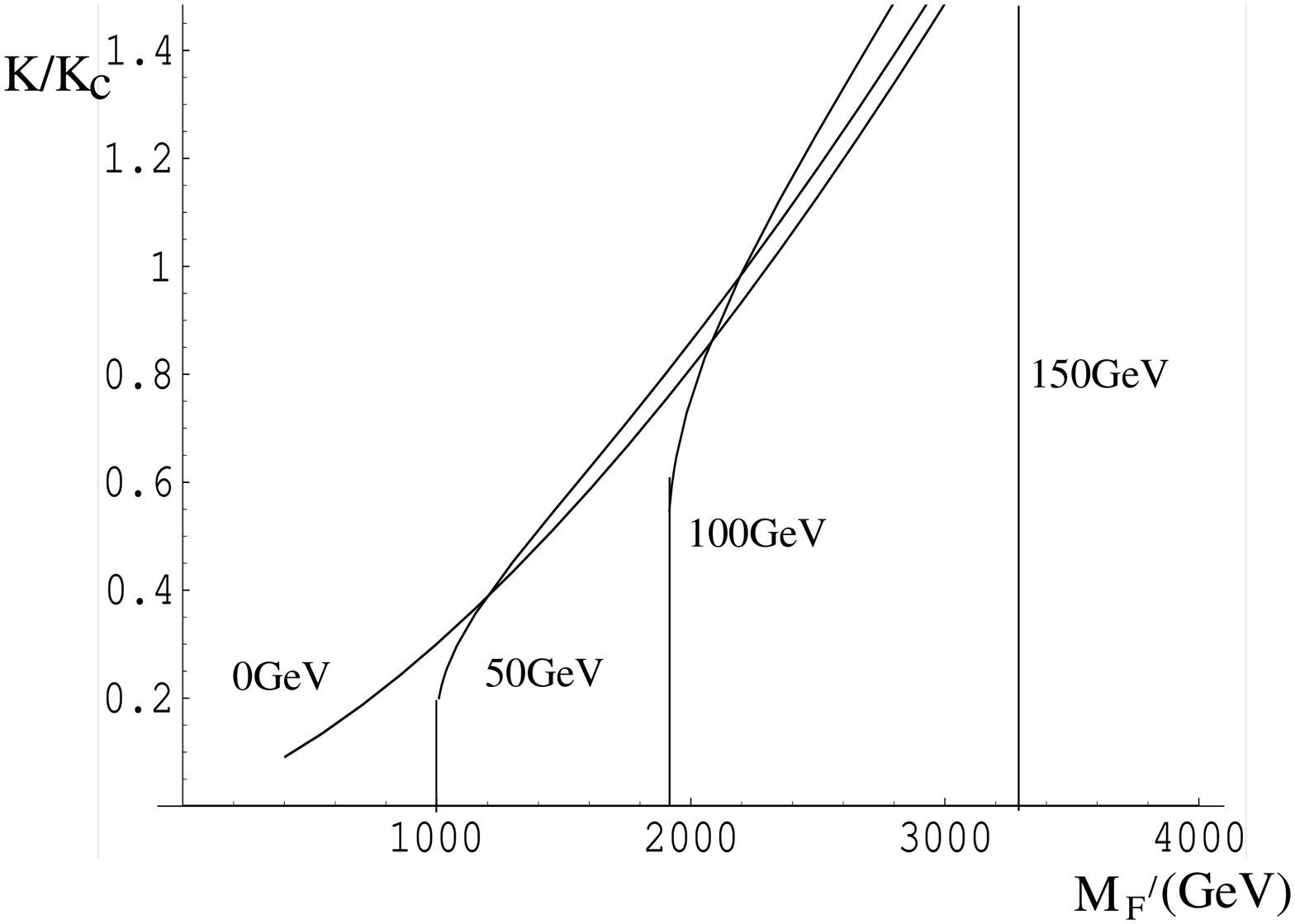}} \vspace{-2cm}
\begin{center}
Figure 15: The $95\%$ exclusion curve in the $\kappa / \kappa_c$-$M_{F}$ 
plane for the gauged chiral, quark, family symmetry model
with universal SM fermion mixing. The four curves correspond to the indicated
values of the mixing mass $m_{mix}$ (\ref{mix}). We take $m_h = 100$ GeV. 
\end{center}

\section{Conclusions}

Gauged flavor symmetries at low energies have been proposed in models of
dynamical electroweak symmetry breaking and fermion mass generation.
These massive flavor gauge bosons give rise to corrections to precisely
measured electroweak quantities. We have performed fits to the $Z$-pole
electroweak data and placed indirect limits on such new physics. In
particular we studied several models from the literature: the universal
coloron; chiral top color; chiral quark family symmetry; SU(9) and
SU(12) chiral flavor symmetry. The $95\%$ confidence limits on the mass
of the heavy gauge bosons for these models, at their critical coupling
for chiral symmetry breaking, typically lie between 1 and 3 TeV. These
limits are more or less insensitive to changes in the higgs mass and the b
quark asymmetry data. We have also discussed the 
inclusion of universal fermion mixing with
heavy EW singlets - this effect dominates the flavoron bounds when the 
mixing mass rises above 150 GeV.  
The precision limits are sufficiently low that
direct searches at the Tevatron would be expected to be competitive or
stronger. Of the models we have studied only the universal coloron's
direct search limits have been computed to this point \cite{bertram} and
as we have seen the direct search limits are superior.  We will explore
the direct search limits in a later paper.

\newpage

%%%%%%%%%%%%%%%%%%%%%%%%%%%%%%%%%%%%%%%%%%%%%%%%%%%%%%%%%%%%%%%%%%%

\centerline{\bf Acknowledgments}

We thank Bogdan Dobrescu, Chris Hill, and Elizabeth Simmons for comments
on the manuscript.  {\em This work was supported in part by the
  Department of Energy under grants DE-FG02-91ER40676 and
  DE-FG02-95ER40896, and by the University of Wisconsin Research
  Committee with funds granted by the Wisconsin Alumni Research
  Foundation.  NE is grateful for the support of a PPARC Advanced
  Research Fellowship.}

%%%%%%%%%%%%%%%%%%%%%%%%%%%%%%%%%%%%%%%%%%%%%%%%%%%%%%%%%%%%%%%%%%%%%

\end{document}